\documentclass[twocolumn,superscriptaddress,amsmath, amssymb, amsfonts,preprintnumbers,aps,prd,longbibliography,nofootinbib]{revtex4-2}

\renewcommand{\sec}[1]{\textit{#1. --- }}
\usepackage{graphicx}% Include figure files
\usepackage{dcolumn}% Align table columns on decimal point
\usepackage{bm}% bold math
 \usepackage{amstext}
 \usepackage{amssymb}
\usepackage{subcaption}
\usepackage[normalem]{ulem}

 \usepackage{amsmath}
 \usepackage{graphicx}
 \usepackage{color}
 \usepackage{bbold}
 \usepackage{delimset} % for brackets
\usepackage[colorlinks=true]{hyperref}
\usepackage{orcidlink}
\usepackage{float}

\newcommand{\tabeq}[2]{ \parbox{#1}{  \be\begin{aligned}#2 \end{aligned} \nonumber \ee }}
\newcommand\II{\mathcal{I}}

\newcommand\Zenodo{\href{https://doi.org/10.5281/zenodo.18340333}{\tt Zenodo} \cite{zenodo} }

\newcommand{\CricArrowRight}[1]{%
    \setlength{\@SizeOfCirc}{\maxof{\widthof{#1}}{\heightof{#1}}}%
    \tikz [x=1.0ex,y=1.0ex,line width=.15ex, draw=blue]%
        \draw [->,anchor=center]%
            node (0,0) {#1}%
            (0,1.2\@SizeOfCirc) arc (85:-240:1.2\@SizeOfCirc);%
}%

\usetikzlibrary{decorations.pathmorphing}
\usetikzlibrary{decorations.markings}
\usetikzlibrary{positioning, shapes, snakes, arrows}
 \usepackage{tikz-feynman}

\tikzset{
	graviton/.style={line width=.8pt, -latex,decorate, decoration={snake, segment length=4pt,amplitude=1pt, pre length=.1cm, post length=.25cm}},
	worldline/.style={gray, line width=1pt},
	worldlineBold/.style={black, line width=.6pt},
        background/.style={black,dotted,line width=1pt},
	zUndirected/.style={line width=1pt},
	zParticle/.style={line width=1pt,postaction={decorate},decoration={markings,mark=at position .6 with {\arrow[#1]{latex}}}},
	zParticleF/.style={line width=1pt,postaction={decorate}},
	cscalar/.style={line width=1pt,postaction={decorate},decoration={markings,mark=at position .6 with {\arrow[#1]{latex}}}},
	cscalar2/.style={line width=1pt,postaction={decorate},decoration={markings,mark=at position .8 with {\arrow[#1]{latex}}}},
	photon/.style={line width =.8pt, decorate, decoration={snake, segment length=3pt, amplitude=1.2pt,  pre length=.1cm, post length=.1cm}},
	 mid arrow/.style={postaction={decorate,decoration={
        markings,
        mark=at position .5 with {\arrow[#1]{latex}}}}} ,
        worlddot/.style={dotted, line width=.8pt},
	worlddot2/.style={dotted, line width=1pt}   }

\DeclareFontFamily{OT1}{pzc}{}
\DeclareFontShape{OT1}{pzc}{m}{it}{<-> s * [1.350] pzcmi7t}{}
\DeclareMathAlphabet{\mathpzc}{OT1}{pzc}{m}{it}

\def\cO{\mathcal{O}}

\def\eps{\epsilon}

\def\d{\mathrm{d}}

\renewcommand{\i}{\ensuremath{\mathrm{i}}}

\newcommand{\iO}{\i 0^{+}}

\def\dd{\delta\!\!\!{}^-\!}

\def\d{\mathrm{d}}
\def\eps{\epsilon}

\def\nn{\nonumber}

\def\eqn#1{Eq.~\eqref{#1}}

\def\Eqn#1{Eq.~\eqref{#1}}

\def\Rcite#1{Ref.~\cite{#1}}
\def\Rcites#1{Refs.~\cite{#1}}
 
\newcommand*\Bell{\ensuremath{\boldsymbol\ell}}

\newcommand{\vev}[1]{\langle #1\rangle}

\newcommand{\widebar}{\overline}

\newcommand{\be}{\begin{equation}}
\newcommand{\ee}{\end{equation}}
\newcommand{\ba}{\begin{align}}
\newcommand{\ea}{\end{align}}
\ifx\genfrac\sdflkaj\else\fi

\newcommand{\mn}{{\mu\nu}}

\newcommand{\pin}{p_{\infty}}

\begin{document}
%/tikzfeynman/warn luatex=false
\preprint{HU-EP-26/04-RTG}

\title{Conservative Black Hole Scattering at  Fifth Post-Minkowskian \\  and
Second Self-Force Order}

\author{Mathias Driesse\,\orcidlink{0000-0002-3983-5852}} 
%\email{mathias.driesse@physik.hu-berlin.de}
\affiliation{%
Institut f\"ur Physik, Humboldt-Universit\"at zu Berlin,
10099 Berlin, Germany
}

\author{Gustav Uhre Jakobsen\,\orcidlink{0000-0001-9743-0442}} 
%\email{gustav.uhre.jakobsen@physik.hu-berlin.de}
\affiliation{%
Institut f\"ur Physik, Humboldt-Universit\"at zu Berlin,
10099 Berlin, Germany
}
\affiliation{Max Planck Institut f\"ur Gravitationsphysik (Albert Einstein Institut), 14476 Potsdam, Germany}

\author{Gustav Mogull\,\orcidlink{0000-0003-3070-5717}}
%\email{gustav.mogull@aei.mpg.de} 
\affiliation{%
Institut f\"ur Physik, Humboldt-Universit\"at zu Berlin,
10099 Berlin, Germany
}
\affiliation{Max Planck Institut f\"ur Gravitationsphysik (Albert Einstein Institut), 14476 Potsdam, Germany}
% \affiliation{Centre for Theoretical Physics, Department of Physics and Astronomy, Queen Mary University of London,  London E1~4NS, United Kingdom}
\affiliation{School of Mathematical Sciences, Queen Mary University of London,  London E1~4NS, United Kingdom}

 \author{Christoph Nega\,\orcidlink{0000-0003-0202-536X}}
%\email{c.nega@tum.de} 
\affiliation{Max Planck Institut f\"ur Gravitationsphysik (Albert Einstein Institut), 14476 Potsdam, Germany}
\affiliation{%
Physik Department, Technische Universit\"at M\"unchen, 85748 Garching, Germany
}
 
 \author{Jan Plefka\,\orcidlink{0000-0003-2883-7825}} 
%\email{jan.plefka@hu-berlin.de}
\affiliation{%
Institut f\"ur Physik, Humboldt-Universit\"at zu Berlin,
10099 Berlin, Germany
}

\author{Benjamin Sauer\,\orcidlink{0000-0002-2071-257X}} 
%\email{benjamin.sauer@hu-berlin.de}
\affiliation{%
Institut f\"ur Physik, Humboldt-Universit\"at zu Berlin,
10099 Berlin, Germany
}

\author{Johann Usovitsch\,\orcidlink{0000-0002-3542-2786}} 
%\email{johann.usovitsch@cern.ch}
\affiliation{%
Institut f\"ur Physik, Humboldt-Universit\"at zu Berlin,
10099 Berlin, Germany
}

\begin{abstract}
Using the worldline quantum field theory formalism, we compute {conservative
contributions to the scattering angle
and impulse} for classical black hole scattering at fifth post-Minkowskian (5PM)
{and second self-force (2SF) order.}
This four-loop calculation involves non-planar Feynman integrals and requires advanced integration-by-parts reduction, novel differential-equation strategies, and efficient boundary-integral algorithms to solve a system of hundreds of master integrals in four integral families on high-performance computing systems.
The resulting function space includes multiple polylogarithms as well as iterated integrals with a K3 period, which generate a spurious velocity divergence at $v/c=\sqrt{8}/3${, $\gamma=3$}. 
This divergence is present in the potential region and must be canceled by contributions from the radiative memory region, while its dimensional-regularisation pole should cancel against the radiative tail region.
{As the standard use of Feynman propagators fails to ensure this cancellation,
we instead propose a ``($\gamma$-3)’’ conservative prescription
that realises both cancellations, leading to a physically sensible answer.}
All available low-velocity checks of our result against the post-Newtonian literature are satisfied.
\end{abstract}
 
\maketitle 

A decade after the first gravitational wave observation emitted by a binary black hole merger \cite{LIGOScientific:2016aoc}, the LIGO–Virgo–KAGRA collaboration \cite{LIGOScientific:2016aoc, LIGOScientific:2017vwq, KAGRA:2021vkt} now reports 218 detections of compact binary coalescences in our universe \cite{LIGOScientific:2025slb}. In the coming decade, a third generation of ground- and space-based gravitational-wave detectors is scheduled to go online \cite{LISA:2017pwj, Punturo:2010zz, Ballmer:2022uxx}, which will dramatically increase the accuracy and frequency range of observations. This will open a new window into gravitational, astrophysical, nuclear, and fundamental physics. To fully exploit this observational potential, theoretical predictions for the dynamics and gravitational radiation of compact binaries must reach a comparable level of precision. This challenge has driven a wide effort comprising the perturbative schemes of post-Newtonian (PN)~\cite{Blanchet:2013haa, Porto:2016pyg, Levi:2018nxp,Brunello:2025gpf}, post-Minkowskian (PM)~\cite{Kosower:2022yvp, Bjerrum-Bohr:2022blt, Buonanno:2022pgc, DiVecchia:2023frv, Jakobsen:2023oow}, and gravitational self-force (SF)~\cite{Mino:1996nk, Poisson:2011nh, Barack:2018yvs, Gralla:2021qaf} expansions, in close synergy with numerical relativity~\cite{Pretorius:2005gq, Boyle:2019kee, Damour:2014afa}. 
Notably, techniques from perturbative quantum field theory (QFT) and effective field theory have gained a leading role in these approaches, enabling increasingly high-order analytic control over the classical two-body problem in general relativity \cite{Driesse:2024feo}.

The PM expansion is a weak-field expansion in powers of Newton’s constant ($G$), being the natural perturbative framework for the unbound scattering of two compact objects — black holes (BHs) or neutron stars (NSs) —  or highly eccentric bound orbits \cite{Kovacs:1978eu, Westpfahl:1979gu, Bel:1981be, Damour:2017zjx, Hopper:2022rwo}. As long as the separation of the two objects ($\sim|b|$) is large compared to their intrinsic sizes ($\sim Gm$), they are captured in an effective worldline theory of massive point particles coupled to gravity \cite{Goldberger:2004jt}. 
The key observables of the change of momentum (known as impulse), the scattering angle, and the far-field waveform have been systematically computed as loop corrections in this classical field theory including spin, tidal effects, and radiation reaction, {see e.g.~\cite{Kalin:2020mvi, Kalin:2020fhe, Kalin:2020lmz, Mogull:2020sak, Jakobsen:2021smu, Dlapa:2021npj, Dlapa:2021vgp, Mougiakakos:2021ckm, Riva:2021vnj, Dlapa:2022lmu, Dlapa:2023hsl, Liu:2021zxr, Mougiakakos:2022sic, Riva:2022fru, Jakobsen:2021lvp, Jakobsen:2021zvh, Jakobsen:2022fcj, Jakobsen:2022zsx, Jakobsen:2022psy, Shi:2021qsb,Bastianelli:2021nbs, Comberiati:2022cpm,Wang:2022ntx,Ben-Shahar:2023djm, Bhattacharyya:2024aeq, Jakobsen:2023ndj, Jakobsen:2023hig, Jakobsen:2023pvx,Caron-Huot:2025tlq,Ivanov:2025ozg,Combaluzier--Szteinsznaider:2025eoc} for worldline} and \cite{Neill:2013wsa, Luna:2017dtq, Kosower:2018adc, Cristofoli:2021vyo, Bjerrum-Bohr:2013bxa, Bjerrum-Bohr:2018xdl, Bern:2019nnu, Bern:2019crd, Bjerrum-Bohr:2021wwt, Cheung:2020gyp, Bjerrum-Bohr:2021din, DiVecchia:2020ymx, DiVecchia:2021bdo, DiVecchia:2021ndb, DiVecchia:2022piu, Heissenberg:2022tsn, Damour:2020tta, Herrmann:2021tct, Damgaard:2019lfh, Damgaard:2019lfh, Damgaard:2021ipf, Damgaard:2023vnx, Aoude:2020onz, AccettulliHuber:2020dal, Brandhuber:2021eyq, Bern:2021dqo, Bern:2021yeh, Bern:2022kto, Bern:2023ity, Damgaard:2023ttc, Brandhuber:2023hhy, Brandhuber:2023hhy, Brandhuber:2023hhl, DeAngelis:2023lvf, Herderschee:2023fxh, Caron-Huot:2023vxl, FebresCordero:2022jts, Bohnenblust:2023qmy} for amplitude based approaches. Both approaches have delivered the scattering angle and impulse up to fourth post-Minkowskian (4PM) order \cite{Bern:2021dqo, Bern:2021yeh, Dlapa:2022lmu, Dlapa:2023hsl, Jakobsen:2023ndj, Jakobsen:2023hig, Jakobsen:2023pvx, Damgaard:2023ttc}.
Employing the worldline quantum field theory (WQFT) formalism \cite{Mogull:2020sak, Jakobsen:2022psy, Jakobsen:2023oow, Haddad:2024ebn}, this progress has culminated in the first determination of the conservative and dissipative fifth post-Minkowskian (5PM) contributions to the scattering angle and impulse at first self-force (1SF) order \cite{Driesse:2024xad, Driesse:2024feo}, i.e.~the leading and sub-leading mass ratio contributions. State-of-the-art numerical computations have recently validated these high-order analytical PM predictions to an impressive degree \cite{Long:2025tvk, Warburton:2025ymy}. 

Despite these advances, the last missing ingredient at 5PM  order has been the sub-sub-leading mass ratio or second self-force (2SF) order contributions to observables. {While 5PM-2SF results for potential graviton modes were established in  $\mathcal{N}=8$ supergravity~\cite{Bern:2025zno}
and very recently in gravity~\cite{Bern:2025wyd} 
in an ordinary differential equation based formal PM expression,
% presented explicitly as a low-velocity expansion,
a complete 5PM-2SF order description has remained out of reach.
Here, we compute conservative contributions to the 5PM-2SF scattering observables using WQFT}.
Our results characterise the {prescription-dependent conservative 5PM scattering dynamics,
providing a crucial step towards} new analytic input for high-accuracy
models tailored to the upcoming generation of gravitational-wave observatories,
{pending the transformation to the bound case \cite{Kalin:2019rwq,Kalin:2019inp,Cho:2021arx,Dlapa:2024cje,Dlapa:2025biy}.}

\begin{figure*}[ht!]
\centering
%%
% (a) 
\begin{subfigure}{0.13\textwidth}
  \centering
  \begin{tikzpicture}[baseline={([yshift=-1ex]current bounding box.south)},scale=.7]
    \coordinate (inA)  at (0.6,0.6);
    \coordinate (outA) at (2.625,0.6);
    \coordinate (inB)  at (0.6,-0.6);
    \coordinate (outB) at (2.625,-0.6);

    \draw[dotted] (inA) -- (outA);
    \draw[dotted] (inB) -- (outB);

    \coordinate (aA) at (0.6,0.6);
    \coordinate (bA) at (1.275,0.6);
    \coordinate (cA) at (1.95,0.6);
    \coordinate (dA) at (2.625,0.6);
    \coordinate (eA) at (3.3,0.6);
    
    \coordinate (aM) at (0.6,0);
    \coordinate (bM) at (1.275,0);
    \coordinate (cM) at (1.95,0);
    \coordinate (dM) at (2.625,0);
    \coordinate (eM) at (3.3,0);

    \coordinate (aB) at (0.6,-0.6);
    \coordinate (bB) at (1.275,-0.6);
    \coordinate (cB) at (1.95,-0.6);
    \coordinate (dB) at (2.625,-0.6);
    \coordinate (eB) at (3.3,-0.6);

    % vertical photons: middle two red
    \draw[photon]     (aA) -- (aB) node [midway,left] {$q\,\,\,\,\,$} node [midway,left] {\mbox{\large$\uparrow$}};
    \draw[photon,red] (aM) -- (dM);
    \draw[photon] (bA) -- (bM);
    \draw[photon] (cB) -- (cM);
    \draw[photon]     (dA) -- (dB);

    % top thick segment only between b and c
%    \draw[zUndirected] (bA) -- (cA);

    % vertices
    \foreach \v in {aA,bA,dA,aM,bM,cM,dM,aB,cB,dB}
      \draw[fill] (\v) circle (.08);
  \end{tikzpicture}
  \caption{}
\end{subfigure}
\hfill
% (b) 
\begin{subfigure}{0.13\textwidth}
  \centering
  \begin{tikzpicture}[baseline={([yshift=-1ex]current bounding box.south)},scale=.7]
     \coordinate (inA)  at (0.6,0.6);
    \coordinate (outA) at (2.625,0.6);
    \coordinate (inB)  at (0.6,-0.6);
    \coordinate (outB) at (2.625,-0.6);

    \draw[dotted] (inA) -- (outA);
    \draw[dotted] (inB) -- (outB);

    \coordinate (aA) at (0.6,0.6);
    \coordinate (bA) at (1.275,0.6);
    \coordinate (cA) at (1.95,0.6);
    \coordinate (dA) at (2.625,0.6);
    \coordinate (eA) at (3.3,0.6);
    
    \coordinate (aM) at (0.6,0);
    \coordinate (bM) at (1.275,0);
    \coordinate (cM) at (1.95,0);
    \coordinate (dM) at (2.625,0);
    \coordinate (eM) at (3.3,0);

    \coordinate (aB) at (0.6,-0.6);
    \coordinate (bB) at (1.275,-0.6);
    \coordinate (cB) at (1.95,-0.6);
    \coordinate (dB) at (2.625,-0.6);
    \coordinate (eB) at (3.3,-0.6);

    % vertical photons: middle two red
    \draw[photon]     (aA) -- (aB);
    \draw[photon,red] (bM) -- (dM);
     \draw[photon,red] (bM) -- (bB);
        \draw[photon] (bA) -- (bM);
    \draw[photon] (cB) -- (cM);
    \draw[photon]     (dA) -- (dB);

    % top thick segment only between b and c
    \draw[zUndirected] (aB) -- (bB);

    % vertices
    \foreach \v in {aA,bA,dA,bB,bM,cM,dM,aB,cB,dB}
      \draw[fill] (\v) circle (.08);
  \end{tikzpicture}
  \caption{}
\end{subfigure}
\hfill
% (c) 
\begin{subfigure}{0.13\textwidth}
  \centering
  \begin{tikzpicture}[baseline={([yshift=-1ex]current bounding box.south)},scale=.7]
     \coordinate (inA)  at (0.6,0.6);
    \coordinate (outA) at (2.625,0.6);
    \coordinate (inB)  at (0.6,-0.6);
    \coordinate (outB) at (2.625,-0.6);

    \draw[dotted] (inA) -- (outA);
    \draw[dotted] (inB) -- (outB);

    \coordinate (aA) at (0.6,0.6);
    \coordinate (bA) at (1.275,0.6);
    \coordinate (cA) at (1.95,0.6);
    \coordinate (dA) at (2.625,0.6);
    \coordinate (eA) at (3.3,0.6);
    
    \coordinate (aM) at (0.6,0);
    \coordinate (bM) at (1.275,0);
    \coordinate (cM) at (1.95,0);
    \coordinate (dM) at (2.625,0);
    \coordinate (eM) at (3.3,0);

    \coordinate (aB) at (0.6,-0.6);
    \coordinate (bB) at (1.275,-0.6);
    \coordinate (cB) at (1.95,-0.6);
    \coordinate (dB) at (2.625,-0.6);
    \coordinate (eB) at (3.3,-0.6);
    
    % vertical photons: middle two red
    \draw[photon]     (aA) -- (aB);
    \draw[photon,red] (cM) -- (dM);
     \draw[photon,red] (cM) -- (cA);
        \draw[photon,red] (bA) -- (bB);
    \draw[photon] (cB) -- (cM);
    \draw[photon]     (dA) -- (dB);

    % top thick segment only between b and c
    \draw[zUndirected] (aB) -- (bB);
      \draw[zUndirected] (bA) -- (cA);

    % vertices
    \foreach \v in {aA,bA,dA,bB,cA,cM,dM,aB,cB,dB}
      \draw[fill] (\v) circle (.08);
  \end{tikzpicture}
  \caption{}
\end{subfigure}
\hfill
% (d) 
\begin{subfigure}{0.13\textwidth}
  \centering
  \begin{tikzpicture}[baseline={([yshift=-1ex]current bounding box.south)},scale=.7]
    \coordinate (inA)  at (0.6,0.6);
    \coordinate (outA) at (2.625,0.6);
    \coordinate (inB)  at (0.6,-0.6);
    \coordinate (outB) at (2.625,-0.6);

    \draw[dotted] (inA) -- (outA);
    \draw[dotted] (inB) -- (outB);

    \coordinate (aA) at (0.6,0.6);
    \coordinate (bA) at (1.275,0.6);
    \coordinate (cA) at (1.95,0.6);
    \coordinate (dA) at (2.625,0.6);
    \coordinate (eA) at (3.3,0.6);
    
    \coordinate (aM) at (0.6,0);
    \coordinate (bM) at (1.275,0);
    \coordinate (cM) at (1.95,0);
    \coordinate (dM) at (2.625,0);
    \coordinate (eM) at (3.3,0);

    \coordinate (aB) at (0.6,-0.6);
    \coordinate (bB) at (1.275,-0.6);
    \coordinate (cB) at (1.95,-0.6);
    \coordinate (dB) at (2.625,-0.6);
    \coordinate (eB) at (3.3,-0.6);

    % vertical photons: middle two red
    \draw[photon]     (aA) -- (aB);
    \draw[photon,red] (bM) -- (cM);
     \draw[photon,red] (cM) -- (cA);
     \draw[photon,red] (bM) -- (bB);
        \draw[photon] (bA) -- (bM);
    \draw[photon] (cB) -- (cM);
    \draw[photon]     (dA) -- (dB);

    % top thick segment only between b and c
    \draw[zUndirected] (aB) -- (bB);
     \draw[zUndirected] (cA) -- (dA);

    % vertices
    \foreach \v in {aA,bA,dA,bB,bM,cM,cA,aB,cB,dB}
      \draw[fill] (\v) circle (.08);
  \end{tikzpicture}
  \caption{}
\end{subfigure}
\hfill
% (e) 
\begin{subfigure}{0.13\textwidth}
  \centering
  \begin{tikzpicture}[baseline={([yshift=-1ex]current bounding box.south)},scale=.7]
    \coordinate (inA)  at (0.6,0.6);
    \coordinate (outA) at (3.3,0.6);
    \coordinate (inB)  at (0.6,-0.6);
    \coordinate (outB) at (3.3,-0.6);

    \draw[dotted] (inA) -- (outA);
    \draw[dotted] (inB) -- (outB);

    \coordinate (aA) at (0.6,0.6);
    \coordinate (bA) at (1.275,0.6);
    \coordinate (cA) at (1.95,0.6);
    \coordinate (dA) at (2.625,0.6);
    \coordinate (eA) at (3.3,0.6);
    
    \coordinate (aB) at (0.6,-0.6);
    \coordinate (bB) at (1.275,-0.6);
    \coordinate (cB) at (1.95,-0.6);
    \coordinate (dB) at (2.625,-0.6);
    \coordinate (eB) at (3.3,-0.6);
    
     % vertical photons: middle two red
    \draw[photon]     (aA) -- (aB);
    \draw[photon]     (bA) -- (bB);
    \draw[photon,red] (cA) -- (cB);
    \draw[photon]	  (dA) -- (dB);
    \draw[photon]     (eA) -- (eB);

    % top thick segment only between b and c
    \draw[zUndirected] (aB) -- (cB);
      \draw[zUndirected] (cA) -- (eA);

    % vertices
    \foreach \v in {aA,bA,cA,dA,eA,aB,bB,cB,dB,eB}
      \draw[fill] (\v) circle (.08);
  \end{tikzpicture}
  \caption{}
\end{subfigure}
\hfill
% (f) 
\begin{subfigure}{0.13\textwidth}
  \centering
  \begin{tikzpicture}[baseline={([yshift=-1ex]current bounding box.south)},scale=.7]
    \coordinate (inA)  at (0.6,0.6);
    \coordinate (outA) at (3.3,0.6);
    \coordinate (inB)  at (0.6,-0.6);
    \coordinate (outB) at (3.3,-0.6);

    \draw[dotted] (inA) -- (outA);
    \draw[dotted] (inB) -- (outB);

    \coordinate (aA) at (0.6,0.6);
    \coordinate (bA) at (1.275,0.6);
    \coordinate (cA) at (1.95,0.6);
    \coordinate (dA) at (2.625,0.6);
    \coordinate (eA) at (3.3,0.6);
    
    \coordinate (aB) at (0.6,-0.6);
    \coordinate (bB) at (1.275,-0.6);
    \coordinate (cB) at (1.95,-0.6);
    \coordinate (dB) at (2.625,-0.6);
    \coordinate (eB) at (3.3,-0.6);
    
     % vertical photons: middle two red
    \draw[photon]     (aA) -- (aB);
    \draw[photon,red]     (bA) -- (bB);
    \draw[photon,red] (cA) -- (cB);
    \draw[photon,red]	  (dA) -- (dB);
    \draw[photon]     (eA) -- (eB);

    % top thick segment only between b and c
      \draw[zUndirected] (bA) -- (cA);
       \draw[zUndirected] (dA) -- (eA);
    \draw[zUndirected] (aB) -- (bB);
     \draw[zUndirected] (cB) -- (dB);

    % vertices
    \foreach \v in {aA,bA,cA,dA,eA,aB,bB,cB,dB,eB}
      \draw[fill] (\v) circle (.08);
  \end{tikzpicture}
  \caption{}
\end{subfigure}
\hfill
% (g) 
\begin{subfigure}{0.13\textwidth}
  \centering
  \begin{tikzpicture}[baseline={([yshift=-1ex]current bounding box.south)},scale=.7]
    \coordinate (inA)  at (0.6,0.6);
    \coordinate (outA) at (3.3,0.6);
    \coordinate (inB)  at (0.6,-0.6);
    \coordinate (outB) at (3.3,-0.6);

    \draw[dotted] (inA) -- (outA);
    \draw[dotted] (inB) -- (outB);

    \coordinate (aA) at (0.6,0.6);
    \coordinate (bA) at (1.275,0.6);
    \coordinate (cA) at (1.95,0.6);
    \coordinate (dA) at (2.625,0.6);
    \coordinate (eA) at (3.3,0.6);
    
    \coordinate (aB) at (0.6,-0.6);
    \coordinate (bB) at (1.275,-0.6);
    \coordinate (cB) at (1.95,-0.6);
    \coordinate (dB) at (2.625,-0.6);
    \coordinate (eB) at (3.3,-0.6);
    
     % vertical photons: middle two red
    \draw[photon]     (aA) -- (aB);
    \draw[photon,red]     (bA) -- (bB);
    \draw[photon] (cA) -- (cB);
    \draw[photon,red]	  (dA) -- (dB);
    \draw[photon]     (eA) -- (eB);

    % top thick segment only between b and c
      \draw[zUndirected] (bA) -- (dA);
      % \draw[zUndirected] (dA) -- (eA);
    \draw[zUndirected] (aB) -- (bB);
     \draw[zUndirected] (dB) -- (eB);

    % vertices
    \foreach \v in {aA,bA,cA,dA,eA,aB,bB,cB,dB,eB}
      \draw[fill] (\v) circle (.08);
  \end{tikzpicture}
  \caption{}
\end{subfigure}
% (h) 
\begin{subfigure}{0.13\textwidth}
  \centering
  \begin{tikzpicture}[baseline={([yshift=-1ex]current bounding box.south)},scale=.7]
    \coordinate (inA)  at (0.6,0.6);
    \coordinate (outA) at (3.3,0.6);
    \coordinate (inB)  at (0.6,-0.6);
    \coordinate (outB) at (3.3,-0.6);

    \draw[dotted] (inA) -- (outA);
    \draw[dotted] (inB) -- (outB);

    \coordinate (aA) at (0.6,0.6);
    \coordinate (bA) at (1.275,0.6);
    \coordinate (cA) at (1.95,0.6);
    \coordinate (dA) at (2.625,0.6);
    \coordinate (eA) at (3.3,0.6);
    
    \coordinate (aB) at (0.6,-0.6);
    \coordinate (bB) at (1.275,-0.6);
    \coordinate (cB) at (1.95,-0.6);
    \coordinate (dB) at (2.625,-0.6);
    \coordinate (eB) at (3.3,-0.6);
    
     % vertical photons: middle two red
    \draw[photon]     (aA) -- (aB) node [midway,left] {$q\,\,\,\,\,$} node [midway,left] {\mbox{\large$\uparrow$}};
    \draw[photon,red] (bA) -- (bB);
    \draw[photon,red] (cA) -- (cB);
    \draw[photon]	  (dA) -- (eB);
    \draw[fill, white] (2.96,0) circle (00.15);
    \draw[photon]     (eA) -- (dB);

    % top thick segment only between b and c
      \draw[zUndirected] (aA) -- (bA);
      \draw[zUndirected] (cA) -- (dA);
      \draw[zUndirected] (bB) -- (dB);
%      \draw[zUndirected] (dB) -- (eB);

    % vertices
    \foreach \v in {aA,bA,cA,dA,eA,aB,bB,cB,dB,eB}
      \draw[fill] (\v) circle (.08);
  \end{tikzpicture}
  \caption{}
\end{subfigure}
% (i) 
\,\,\,\,\,\,\,\,\,\,\,\,\,\,
\begin{subfigure}{0.13\textwidth}
  \centering
  \begin{tikzpicture}[baseline={([yshift=-1ex]current bounding box.south)},scale=.7]
    \coordinate (inA)  at (0.6,0.6);
    \coordinate (outA) at (3.3,0.6);
    \coordinate (inB)  at (0.6,-0.6);
    \coordinate (outB) at (3.3,-0.6);

    \draw[dotted] (inA) -- (outA);
    \draw[dotted] (inB) -- (outB);

    \coordinate (aA) at (0.6,0.6);
    \coordinate (bA) at (1.275,0.6);
    \coordinate (cA) at (1.95,0.6);
    \coordinate (dA) at (2.625,0.6);
    \coordinate (eA) at (3.3,0.6);
    
    \coordinate (aB) at (0.6,-0.6);
    \coordinate (bB) at (1.275,-0.6);
    \coordinate (cB) at (1.95,-0.6);
    \coordinate (dB) at (2.625,-0.6);
    \coordinate (eB) at (3.3,-0.6);
    
     % vertical photons: middle two red
    \draw[photon]     (aA) -- (aB);
    \draw[photon,red] (bA) -- (cB);
    \draw[fill, white] (1.61,0) circle (00.15);
    \draw[photon] (cA) -- (bB);
    \draw[photon]	  (dA) -- (dB);
   % \draw[fill, white] (2.96,0) circle (00.15);
    \draw[photon]     (eA) -- (eB);

    % top thick segment only between b and c
      \draw[zUndirected] (aA) -- (cA);
    %  \draw[zUndirected] (cA) -- (dA);
      \draw[zUndirected] (cB) -- (eB);
%      \draw[zUndirected] (dB) -- (eB);

    % vertices
    \foreach \v in {aA,bA,cA,dA,eA,aB,bB,cB,dB,eB}
      \draw[fill] (\v) circle (.08);
  \end{tikzpicture}
  \caption{}
\end{subfigure}
\hfill
% (j) 
\begin{subfigure}{0.13\textwidth}
  \centering
  \begin{tikzpicture}[baseline={([yshift=-1ex]current bounding box.south)},scale=.7]
    \coordinate (inA)  at (0.6,0.6);
    \coordinate (outA) at (3.3,0.6);
    \coordinate (inB)  at (0.6,-0.6);
    \coordinate (outB) at (3.3,-0.6);

    \draw[dotted] (inA) -- (outA);
    \draw[dotted] (inB) -- (outB);

    \coordinate (aA) at (0.6,0.6);
    \coordinate (bA) at (1.275,0.6);
    \coordinate (cA) at (1.95,0.6);
    \coordinate (dA) at (2.625,0.6);
    \coordinate (eA) at (3.3,0.6);
    
    \coordinate (aB) at (0.6,-0.6);
    \coordinate (bB) at (1.275,-0.6);
    \coordinate (cB) at (1.95,-0.6);
    \coordinate (dB) at (2.625,-0.6);
    \coordinate (eB) at (3.3,-0.6);
    
     % vertical photons: middle two red
    \draw[photon]     (aA) -- (bB);
    \draw[fill, white] (0.94,0) circle (00.15);
    \draw[photon] (bA) -- (aB);
    \draw[photon,red] (cA) -- (cB);
    \draw[photon]	  (dA) -- (eB);
    \draw[fill, white] (2.96,0) circle (00.15);
    \draw[photon]     (eA) -- (dB);

    % top thick segment only between b and c
      \draw[zUndirected] (bA) -- (dA);
    %  \draw[zUndirected] (cA) -- (dA);
      \draw[zUndirected] (bB) -- (dB);
%      \draw[zUndirected] (dB) -- (eB);

    % vertices
    \foreach \v in {aA,bA,cA,dA,eA,aB,bB,cB,dB,eB}
      \draw[fill] (\v) circle (.08);
  \end{tikzpicture}
  \caption{}
\end{subfigure}
\hfill
% (k) 
\begin{subfigure}{0.13\textwidth}
  \centering
\begin{tikzpicture}[baseline={([yshift=-1ex]current bounding box.south)},scale=.7]
  % External (dotted) lines
  
   \coordinate (inA)  at (0.6,0.6);
    \coordinate (outA) at (2.625,0.6);
    \coordinate (inB)  at (0.6,-0.6);
    \coordinate (outB) at (2.625,-0.6);

    \draw[dotted] (inA) -- (outA);
    \draw[dotted] (inB) -- (outB);

    \coordinate (aA) at (0.6,0.6);
    \coordinate (bA) at (1.85+0.1,0.6);
    \coordinate (cA) at (1.95,0.6);
    \coordinate (dA) at (2.625,0.6);
    \coordinate (eA) at (3.3,0.6);
    
    \coordinate (aM) at (0.6,0);
    \coordinate (bM) at (1.275,0);
    \coordinate (cM) at (1.95,0);
    \coordinate (dM) at (2.625,0);
    \coordinate (eM) at (3.3,0);

    \coordinate (aB) at (0.6,-0.6);
    \coordinate (bB) at (1.275,-0.6);
    \coordinate (cB) at (1.95,-0.6);
    \coordinate (dB) at (2.625,-0.6);
    \coordinate (eB) at (3.3,-0.6);
      
  % Extra vertex on central vertical line (for tadpole and jump)
  \coordinate (bL) at (1.85+0.1,-0.35);

  % Horizontal active propagator (red) in the middle
  \draw[photon,red] (aM) -- (dM);

  \draw[fill, white] (1.85+0.1,0) circle (0.15);
  
  % Vertical photons (black)
  \draw[photon] (aA) -- (aB);
 \draw[photon,red] (bM) -- (bL);
  \draw[photon] (bL) -- (bB);
  \draw[photon] (dA) -- (dB);
   \draw[photon] (bL) -- (bA);
        
         % Non-planar jumping active propagator (red) from yL to yA
 % \draw[photon] (bL) .. controls (1.9+0.2,0.1) and (1.9+0.2,0.1) .. (bA);

  % Filled vertices
  \foreach \v in {aA,bA,dA,aM,bM,dM,aB,bB,dB,bL}
    \draw[fill] (\v) circle (.08);
\end{tikzpicture}
  \caption{}
\end{subfigure}
% (l) 
\begin{subfigure}{0.13\textwidth}
  \centering
\begin{tikzpicture}[baseline={([yshift=-1ex]current bounding box.south)},scale=.7]
  % External (dotted) lines
  
   \coordinate (inA)  at (0.6,0.6);
    \coordinate (outA) at (2.625,0.6);
    \coordinate (inB)  at (0.6,-0.6);
    \coordinate (outB) at (2.625,-0.6);

    \draw[dotted] (inA) -- (outA);
    \draw[dotted] (inB) -- (outB);

    \coordinate (aA) at (0.6,0.6);
    \coordinate (bA) at (1.275,0.6);
    \coordinate (cA) at (1.95,0.6);
    \coordinate (dA) at (2.625,0.6);
    \coordinate (eA) at (3.3,0.6);
    
    \coordinate (aM) at (0.6,0);
    \coordinate (bM) at (1.275,0);
    \coordinate (cM) at (1.95,0);
    \coordinate (dM) at (2.625,0);
    \coordinate (eM) at (3.3,0);

    \coordinate (aB) at (0.6,-0.6);
    \coordinate (bB) at (1.275,-0.6);
    \coordinate (cB) at (1.95,-0.6);
    \coordinate (dB) at (2.625,-0.6);
    \coordinate (eB) at (3.3,-0.6);
      
  % Extra vertex on central vertical line (for tadpole and jump)
  %\coordinate (bL) at (1.95,0.35);

  % Horizontal active propagator (red) in the middle
  \draw[photon,red] (aM) -- (dM);

  \draw[fill, white] (1.275,0) circle (0.15);

  % Vertical photons (black)
  \draw[photon] (aA) -- (aB);
 \draw[photon] (bA) -- (bB);
 \draw[photon, red] (cM) -- (cB);
  \draw[photon] (dA) -- (dB);
%  \draw[photon] (bA) -- (bL);
        
         % Non-planar jumping active propagator (red) from yL to yA
% \draw[photon] (bL) .. controls (1.9,0.1) and (1.9,0.1) .. (cB);

 \draw[zUndirected] (bB) -- (cB);
  % Filled vertices
  \foreach \v in {aA,bA,aB,bB,cB,dB,dA,aM,dM,cM}
    \draw[fill] (\v) circle (.08);
\end{tikzpicture}
  \caption{}
\end{subfigure}
% (m) 
\begin{subfigure}{0.13\textwidth}
  \centering
\begin{tikzpicture}[baseline={([yshift=-1ex]current bounding box.south)},scale=.7]
  % External (dotted) lines
  
   \coordinate (inA)  at (0.6,0.6);
    \coordinate (outA) at (2.625,0.6);
    \coordinate (inB)  at (0.6,-0.6);
    \coordinate (outB) at (2.625,-0.6);

    \draw[dotted] (inA) -- (outA);
    \draw[dotted] (inB) -- (outB);

    \coordinate (aA) at (0.6,0.6);
    \coordinate (bA) at (1.275,0.6);
    \coordinate (cA) at (1.95,0.6);
    \coordinate (dA) at (2.625,0.6);
    \coordinate (eA) at (3.3,0.6);
    
    \coordinate (aM) at (0.6,0);
    \coordinate (bM) at (1.275,0);
    \coordinate (cM) at (1.95,0);
    \coordinate (dM) at (2.625,0);
    \coordinate (eM) at (3.3,0);

    \coordinate (aB) at (0.6,-0.6);
    \coordinate (bB) at (1.275,-0.6);
    \coordinate (cB) at (1.95,-0.6);
    \coordinate (dB) at (2.625,-0.6);
    \coordinate (eB) at (3.3,-0.6);
      
  % Extra vertex on central vertical line (for tadpole and jump)
  \coordinate (bL) at (1.95,0.35);

  % Horizontal active propagator (red) in the middle
  \draw[photon] (aB) -- (bA);

  \draw[fill, white] (0.95,0) circle (0.15);

  % Vertical photons (black)
  \draw[photon] (aA) -- (bB);
 \draw[photon,red] (cA) -- (cB);
 \draw[photon, red] (cM) -- (dM);
  \draw[photon] (dA) -- (dB);
  \draw[zUndirected] (bA) -- (cA);      
   \draw[zUndirected] (bB) -- (cB);      
         % Non-planar jumping active propagator (red) from yL to yA
%  \draw[photon] (bL) .. controls (1.9,0.1) and (1.9,0.1) .. (cB);

  % Filled vertices
  \foreach \v in {aA,bA,cB,dA,cM,dM,aB,bB,dB,cA}
    \draw[fill] (\v) circle (.08);
\end{tikzpicture}
  \caption{}
\end{subfigure}
\caption{\small
{The 14 top-level sectors of the four-loop 2SF integral families:
planar P in the first row (a--g);
in the second row nonplanar with one graviton crossing NP1 (h--i),
nonplanar with two graviton crossings NP2 (j),
and nonplanars arising from non-linear memory NPM (k--m),
with an active 3-graviton vertex.
All integrals encountered in our calculation involve subsets of
those appearing in one of these sectors.}
Solid lines denote worldline propagators, {$\ell\cdot v_i$},
wavy lines graviton propagators, {$\ell^2$},
and the dotted lines may be interpreted as a cut worldline propagator, $\dd(\ell\cdot v_{i})$.
In that sense, the top-level sectors have 13 propagators,
and red graviton propagators may go on-shell.
Finally, $q^{\mu}$ is the total momentum transfer.}
\label{fig:diagrams}
\end{figure*}

%===========================
\sec{Worldline Quantum Field Theory}  
The two (non-spinning) BHs or NSs are modelled as point particles with trajectories
$x_{i}^{\mu}(\tau)$.
% finite-size effects being suppressed to higher PM orders.
In proper time gauge $\dot x_{i}^{2}=1$,
\begin{align}\label{eq:action}
S=-\sum_{i=1}^{2}\frac{m_i}2\int\!\d\tau\,g_{\mu\nu}\dot x_i^{\mu}\dot x_i^{\nu}
 - \frac{1}{16\pi G}\int\!\d^{D}x \sqrt{-g} R \, ,
\end{align}
suppressing a gauge-fixing term $S_{\text{gf}}$.
We use a mostly minus signature, dimensional regularisation with $D=4-2\epsilon$ and consider an expansion around the Minkowskian ($G^{0}$) background and straight-line trajectories:
\begin{align}\label{backgroundexp}
    x_i^\mu(\tau) = b_i^\mu \!+\! v_i^\mu \tau \!+\! z_i^\mu(\tau)\,,\,\,
    g_{\mu\nu} = \eta_{\mu\nu} +\sqrt{32\pi G} h_\mn \,,
\end{align}
with the  worldline deflections $z_i^\mu(\tau)$ and graviton field $h_\mn(x)$.
The initial data consists of the impact parameter $b^{\mu}=b_{2}^{\mu}-b_{1}^{\mu}$ and initial velocities $v^\mu_{1},v^\mu_{2}$, with $b\cdot v_i=0$, $v_i^2=1$ and $\gamma=v_1\cdot v_2=(1-v^{2})^{-1/2}$. We also introduce the parameter $x\in[0,1]$ defined via $\gamma=(x+x^{-1})/2$ to rationalise expressions. 

In the WQFT formalism, solutions to the classical equations of motion for the trajectory and metric are given by their tree-level one-point functions,
$\vev{x^{\mu}_{i}(\tau)}$ and $\vev{g_{\mu\nu}(x)}$.
Nevertheless, non-trivial Feynman loop integrals arise due to the hybrid nature of the WQFT:
worldlines conserve only a single momentum component, as opposed to full $D$-dimensional momentum conservation in the bulk.
The impulse of the first BH, $\Delta p_{1}^{\mu}$, emerges from $\Delta p_{1}^{\mu} = \lim_{\omega\to 0}\omega^{2}\vev{z_{1}^{\mu}(\omega)}$ in momentum (energy) space.
At 5PM order, the impulse factorises into SF sectors:
\begin{align}\label{eq:sfExpansion}
\Delta p^{(5) \mu}_1=  m_{1} m_{2} & \Bigl ( m_2^{4} \Delta p^{(5) \mu}_{\text{0SF}}
+  m_1 m_2^3 \Delta p^{(5) \mu}_{\text{1SF}}   \\
+m_1^2 m_2^2 &\Delta p^{(5) \mu}_{\text{2SF}} 
+ m_1^3 m_2 \Delta p^{(5) \mu}_{\widebar{\text{1SF}}} 
 + m_1^4 \Delta p^{(5) \mu}_{\widebar{\text{0SF}}}\Bigr) \nn \,.
\end{align}
In this Letter we compute the unknown even-in-velocity conservative part of $\Delta p^{(5) \mu}_{\text{2SF}}$, while all other parts are known \cite{Driesse:2024xad,Driesse:2024feo}.
On a basis of $b^\mu$, $v_1^\mu$ and $v_2^\mu$ in the scattering plane, we compute the $b$-component directly ---
$v_i$-components are reconstructed using the on-shell conditions $p_i^2=(p_i+\Delta p_i)^2$ and momentum conservation $\Delta p_1^\mu=-\Delta p_2^\mu$.

%===========================
\sec{Integrand}
Generation of the WQFT integrand employs recursive diagrammatic techniques~\cite{Jakobsen:2022fcj, Jakobsen:2022psy, Jakobsen:2022zsx, Jakobsen:2023oow, Jakobsen:2023ndj, Jakobsen:2023hig, Driesse:2024xad, Driesse:2024feo},
producing a total of {651} Feynman diagrams.
At the 5PM order, we insert up to six-graviton vertices, whose complexity 
is reduced 
through an optimised non-linear De Donder gauge~\cite{Driesse:2024feo}.
A subtle point is how to define a \emph{conservative} sector of the truly dissipative
(in-in) problem in which $\Delta p_{1}^{\mu} + \Delta p_{2}^{\mu}\neq 0$ due to momentum 
(and angular momentum) being carried away by gravitational radiation.
This {has previously been} achieved \cite{Kalin:2020fhe, Kalin:2020lmz, Dlapa:2021vgp, Jakobsen:2022fcj, Jakobsen:2023ndj, Jakobsen:2023pvx, Driesse:2024xad} by employing Feynman propagators  (in-out) for the gravitons
and retarded propagators on the worldline (in-in) \cite{Jakobsen:2022psy, Kalin:2022hph}, taking the part real and even in velocity $v$ in the end.
{In this work we take a different approach, discussed as part of our boundary fixing ---
for now, we assume only that graviton propagators are individually time-symmetric.}

% \jan{Change:We initially also follow this prescription, which
% amounts to projecting out certain parts of the integrals.}

Our diagrams organise themselves into 14 top-level sectors with respect to the integration-by-parts (IBP) identities, displayed in Fig.~\ref{fig:diagrams} --- these are indicative of the Feynman diagrams encountered at 2SF.
A vital difference between 1SF and 2SF is that we cannot use partial fraction identities on worldline propagators to ``planarise'' the entire integrand~\cite{Driesse:2024xad}.
Nevertheless, partial fractions are still used to resolve linear identities between worldline propagators.
We require four separate integral families {(P, NP1, NP2, NPM)}, explicit expressions for which are provided in the Supplemental Material.
The planar family (P) accounts for top-sectors (a--g) of Fig.~\ref{fig:diagrams},
{and all diagrams take a ``Mondrian'' structure without lines crossing each other.
The first two nonplanar families (NP1, NP2)} capture the possibility of one
(h--i) or two (j) crossings of the graviton propagators.

{The last non-planar family (NPM),}
associated with top-sectors (k--m), captures the effect of \emph{non-linear memory}~\cite{PhysRevD.46.4304, Porto:2024cwd, Cunningham:2024dog, Georgoudis:2025vkk}.
Non-planarity of these diagrams is carried by a vertex where three \emph{active} gravitons (highlighted in red) meet.
The effect is to produce non-linear corrections to the Einstein equation expanded around a Schwarzschild background.
There is an $S_{3}$ permutation symmetry about this vertex, which we exploit next to
shifts on the loop momenta in order to maximally simplify our integrand as preparation for IBP reduction.

%===========================
\sec{Integration-by-parts reduction} 
{The first step towards full integration was finding linear
IBP identities between the elements of our four integral families,
enabling reduction to a much smaller basis of master integrals.}
Our IBP reduction was performed using
{\tt KIRA 3.0}~\cite{Maierhofer:2017gsa,Klappert:2020nbg, Lange:2021edb, Lange:2025fba}.
This was the computational bottleneck of the calculation,
consuming $\sim3\times10^6$ core hours on a high-performance cluster ---
an order of magnitude larger than 1SF~\cite{Driesse:2024feo, Driesse:2024xad}.
Yet this number masks the true complexity of the 2SF problem,
which required two key additional technical improvements:
\begin{enumerate}
  \item The use of symmetry relations in the IBP reduction specific to the conservative setup led to a reduction in the number of master integrals. For the { P, NP1, and NP2 families}, we let {\tt KIRA} generate these symmetry relations automatically.
  Symmetries of the {NPM} family were generated by hand and given as extra equations.
  \item A careful choice of basis sped up the reduction by an order of magnitude, thereby reducing the polynomial degree of the result.
  This ``pre-canonical" basis is chosen as close to the canonical integrals as possible without introducing any algebraic or transcendental functions.
  When using finite-field reconstruction with {\tt FireFly}~\cite{Klappert:2019emp, Klappert:2020aqs}, we observed a drastic reduction not only in the number of probes per prime field but also in the number of prime fields.
\end{enumerate}
In summary, we find the following numbers of master integrals (MIs) in our four families
\be
\begin{tabular}{l|c|c|c|c}
Family & P & NP1 & NP2 & NPM \cr
\hline
\# MIs & 321 & 144 & 46 & 220
\end{tabular}
\ee

%===========================
\sec{Differential equations}
We compute the master integrals $\underline I(x;\epsilon)=(I_1,\hdots, I_N)$ using differential equations (DEs)~\cite{Kotikov:1990kg, Remiddi:1997ny, Gehrmann:1999as} in the kinematic variable $x$, i.e., $\partial_x \underline I(x;\epsilon) = B(x;\epsilon) \underline I(x;\epsilon) $, where the matrix $B(x;\epsilon)$  is derived using the IBP relations. To systematically calculate the expansion of the master integrals in the dimensional regulator $\epsilon$, we transform the differential equation into canonical form~\cite{Henn:2013pwa}, which is achieved by the rotation $\underline J(x;\epsilon) = T(x;\epsilon)\underline I(x;\epsilon) $. In the new basis, $\epsilon$ factorises in the differential equation
\begin{equation}
	\frac{\partial}{\partial x} \underline J(x;\epsilon)  = \epsilon A(x) \underline J(x;\epsilon) \, ,
\label{eq:can}
\end{equation}
with the transformed matrix $\epsilon A(x) = T(B T^{-1} - \partial_x T^{-1})$. The structure of Eq.~\eqref{eq:can} enables us to write its solution in terms of iterated integrals.

\begin{figure}[t] 
	\includegraphics[width=0.9\hsize]{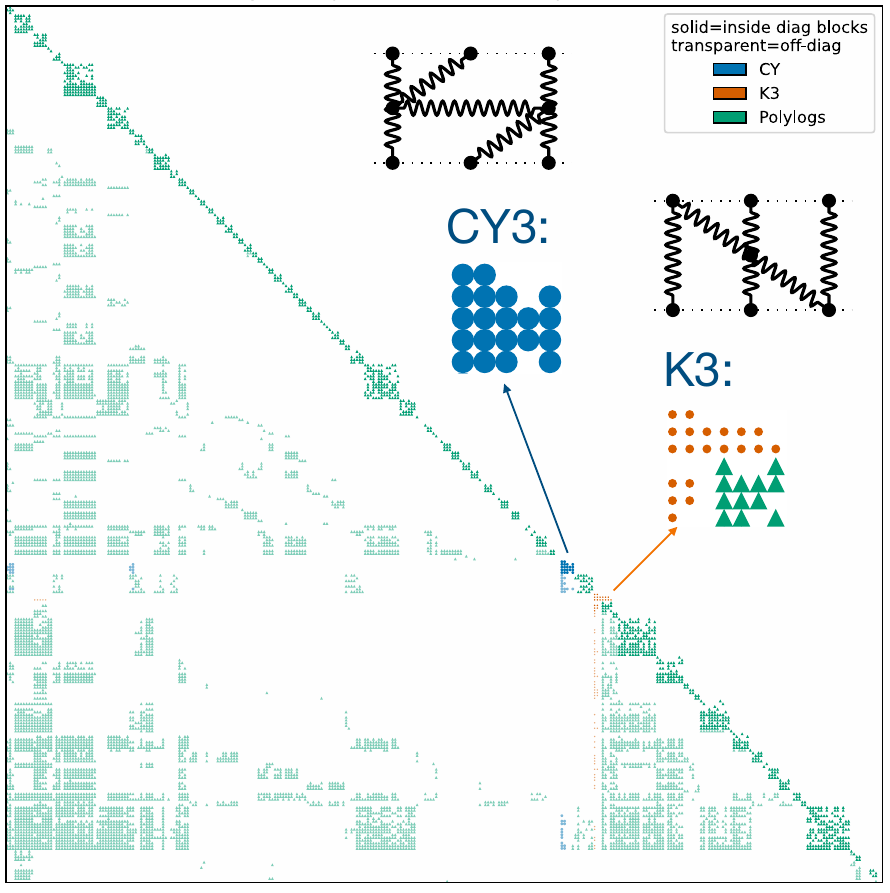}
	 \caption{Non-zero entries of the $321\times 321$ DE matrix 
	 $A(x)$ of the planar family (P). The solid blocks on the diagonals
	 determine the function spaces of CY3 (blue), K3 (orange), and %multi-
	 polylogarithmic (green) type. %The off-diagonal entries are in light-green. 
	 The off-diagonal entries are in corresponding lighter colours.
   %The biggest diagonal block is $16\times 16$.
	}
	  \label{fig-2}
	  \end{figure}

To find the canonical basis $\underline J(x;\epsilon) $, we use the method developed in \Rcites{Gorges:2023zgv, Duhr:2025lbz}. It is essential to understand the different geometries appearing in the individual sectors (groups of integrals that share the same propagators) of the set of master integrals. Two- and three-dimensional Calabi-Yau (CY) varieties are known to appear at the 5PM order~\cite{Frellesvig:2023bbf, Klemm:2024wtd, Frellesvig:2024zph, Driesse:2024feo, Frellesvig:2024rea}. These geometries are natural generalisations of elliptic curves to higher dimensions, and solve Einstein's equations in the vacuum~\cite{Yau:1978cfy}. Traditionally, they have been used in the context of string theory compactifications~\cite{Candelas:1985en, Klemm}, but have recently found applications in Feynman integrals for particle physics~\cite{Vanhove:2014wqa, Bourjaily:2018yfy, Klemm:2019dbm, Bonisch:2020qmm, Bonisch:2021yfw, Bourjaily:2022bwx, Duhr:2022pch}. In our 5PM-2SF problem, we find two different Calabi-Yau varieties of dimension three and two, where the two-dimensional one is also called a K3 surface~\cite{MR3586372} --- see~Fig.~\ref{fig-2}. An analysis of these CY varieties, in particular their definitions via polynomial equations, has already been done by some of the present authors~\cite{Klemm:2024wtd}. The periods $\varpi(x)$ of the CYs are determined by their Picard-Fuchs DEs $\mathcal{L} \varpi(x) = 0$. The corresponding operator for the K3 is given by
($\theta = x\, \partial_x$)
\begin{align}\label{eq:pfk3}
\begin{aligned}
	\mathcal{L}_\text{K3} 	&= \left(1-34 x^2+x^4\right)\theta ^3-6 x^2 \left(17-x^2\right)\theta ^2 \\
						&\quad -12 (3-x) (3+x) x^2 \theta -8 \left(5-x^2\right) x^2	\, , 
\end{aligned}
\end{align}
and is related to the Ap\'ery operator \cite{MR3363457,MR3890449}.
The $\epsilon$-factorisation of the pure CY3 and K3 sectors has been given in \Rcite{Duhr:2025lbz}. 

To compute a fully canonical system, we follow a similar strategy as outlined in our 5PM-1SF computation \cite{Driesse:2024feo}. The increase in complexity of DEs compared with 1SF leads to increased block sizes up to $16\times 16$, c.f.~Fig.~\ref{fig-2}.
For the smaller polylogarithmic diagonal blocks, we employ \texttt{CANONICA}~\cite{Meyer:2017joq}.
For larger sectors, we use the Baikov representation of our integrals \cite{Frellesvig:2024ymq} to find a single integral of uniform transcendental weight (UT). Then we use \texttt{INITIAL}~\cite{Dlapa:2020cwj} to transform the whole diagonal block to canonical form.
To transform the off-diagonal blocks,
we use in-house code based on \texttt{FiniteFlow}~\cite{Peraro:2019svx}
and \texttt{MultivariateApart}~\cite{Heller:2021qkz}.
For the couplings between polylogarithmic sectors and the CY sectors, we employ integrand analysis using the Baikov representation~\cite{Frellesvig:2024ymq} and IBP-based searches for near-canonical integrals. Using successive transformations, these integrals are rotated into canonical form. 

Interestingly, after IBP reduction we find that
only the K3 sector contributes to our final result.
The $\mathcal{L}_\text{K3}$ operator (\ref{eq:pfk3}),
and therefore the integrals in this sector, have a singular point at $x=3-2 \sqrt{2}$,
or equivalently $\gamma=3$, $v=\sqrt{8}/3$, which lies in the physical region $x \in [0, 1]$.
We also needed to introduce the square root $r(x)=\sqrt{-1+34x^2-x^4}$ in the canonicalisation process, which shares the same singularity structure. Singularities at $\gamma=3$ appear in partial results but must cancel in the observables.

%===========================
\sec{Expansion by regions and divergences}
After $\epsilon$-factorization we construct the solution for the master integrals 
in an $\epsilon$ expansion up to boundary integrals. We determine the boundaries in the small velocity limit ($x \rightarrow 1, \gamma\rightarrow1, v\rightarrow 0$) using the  \emph{method of regions} \cite{Beneke:1997zp, Smirnov:2012gma, Becher:2014oda}. 
%This split into regions is also necessary to separate the conservative and dissipative parts of the answer. 
  This split into regions also plays a role in the separation of conservative and dissipative effects.
  The regions are characterised by scalings of loop momenta:
\begin{equation}\label{PRregions}
	  \ell^\text{P}=(\ell^0,\Bell)\sim(v,1)\,,  \qquad
 	  \ell^\text{R}=(\ell^0,\Bell)\sim(v,v) \, .
\end{equation}
We call these \emph{potential} (P) or \emph{radiative} (R) scalings respectively.
Generally, all loop momenta may be potential but only a few may also become radiative ---
the active gravitons, highlighted in red in Fig.~\ref{fig:diagrams}.

\begin{figure}[t] 
	\includegraphics[width=0.9\hsize]{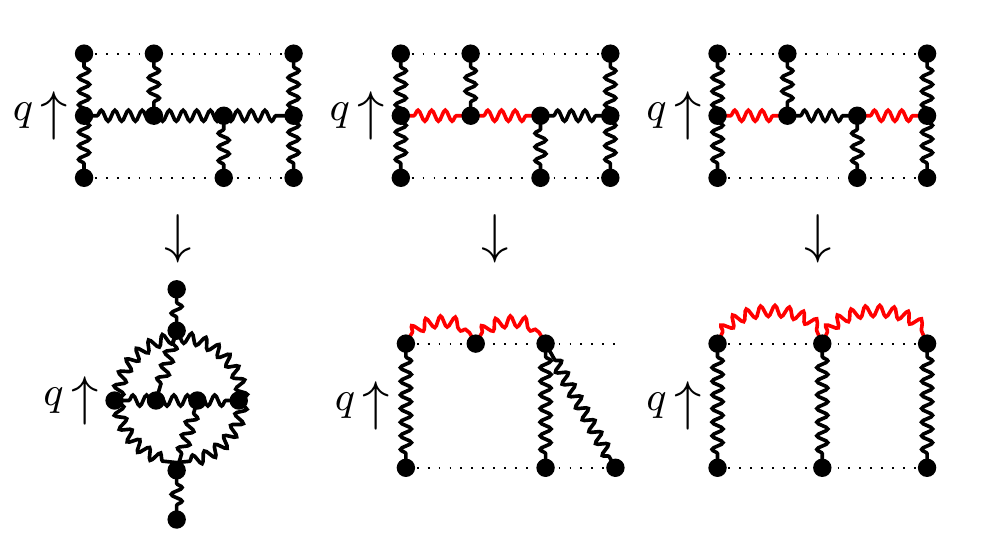}
	\centerline{(a) Potential \hspace{0.9cm} (b) Tail \hspace{0.9cm} (c) Memory }
	 \caption{ Active gravitons that become radiative (red) determine
  the three regions for a planar topsector and yield the required boundary integrals.
  In the tail and memory region, the $\iO$ prescription is crucial.
 }
  \label{fig:boundary}
  	  \end{figure}

{In the first three families (P, NP1 and NP2)} there are in total four propagators $D_{13}, D_{23}, D_{14}, D_{24}$ (defined in  \Eqn{DIJs} of the supplementary material) with active momenta although at most three can be present in the same diagram.
In this work we include conservative contributions from regions with an even number of radiative gravitons (as was done at lower orders \cite{Kalin:2020fhe,Dlapa:2021vgp,Jakobsen:2022fcj,Jakobsen:2023ndj,Driesse:2024xad}).
{There are three regions:}
the potential region PPPP, the \textit{tail} region composed of RRPP, PPRR, RPRP and PRPR and finally the \textit{memory} region composed of RPPR and PRRP which is a novel 2SF feature --- see Fig.~\ref{fig:boundary}.
We use the names \textit{tail} and \textit{memory} as the corresponding regions include the tail effect~\cite{PhysRevD.37.1410, PhysRevD.46.4304, Porto:2024cwd} and memory effect respectively.
The two RR regions, tail and memory, are not related by symmetries in contrast to their subregions.

For the {non-planar memory family NPM},
the momenta of the propagators {$D_5^\text{NPM}, D_6^\text{NPM}, D_7^\text{NPM}$}
(defined in \eqref{DNPs} of the supplementary material) are active.
Only two of those three momenta are, however, linearly independent and give rise to two regions even in R --- PP and RR --- from which we include conservative contributions.
The non-planar RR region is of the memory type (and does not include any tail).

As is known from the 4PM and 5PM-1SF orders, there is an intricate interplay between the various regions in the form of a cancellation of poles in the dimensional regulator $\epsilon$,
leading to terms $\propto \log v$ in the final finite result. 
Here, at our 5PM-2SF order a crucial novel feature arises:
the potential region carries not only an $\epsilon$ pole, but also terms that diverge at $\gamma=3$, which stem from the integrals associated with the K3 geometry.
Such a singularity was also observed in the PN velocity expansion of the potential region in \cite{Bern:2025zno, Bern:2025wyd}, which the authors attributed to the $\gamma=-1$ pole.
This novel $\gamma=3$ divergence does \emph{not} occur in the tail region, while the memory
region carries these velocity divergencies as well. Hence, in order to obtain
a physically meaningful result we need a second intricate interplay between the regions to cancel divergencies.

%===========================

%===========================
\sec{Boundary integrals and the $\iO$ prescription}
The final step is to perform the boundary integrals in the discussed regions. It is here where the crucial $\iO$ prescription for the graviton propagators makes its impact (next to assuming related propagator symmetries in the IBP reduction). We perform the required boundary integrals analytically at the static point $x=1$. Depending on the candidate we use three strategies: (i) 
{ by transforming the integrals into a Feynman parameter representation, where direct integration in terms of hypergeometric functions in closed $\epsilon$ form is possible \cite{Driesse:2024feo}, (ii) by going to time \cite{Driesse:2024feo} or frequency domain representations (see Supplementary Material)} and (iii) by the ``Feynman parameter integration through differential equations" strategy (FP) of \cite{Hidding:2022ycg} to compute the missing integrals numerically. From high-precision numerical data, we utilise the PSLQ algorithm \cite{Ferguson1999PSLQ} to reconstruct the analytic results. 

The commonly used prescription for the conservative sector is Feynman $\iO$ on the gravitons, together with taking the real and even part in $v$ in the final result. The latter is equivalent to only picking an even number of R regions for the boundaries. This prescription led to a cancellation of $\epsilon$-poles between the regions and also matched the conservative results of the PN expansion in the low velocity limits at 3PM and 4PM orders. Yet, already at 5PM-1SF order \cite{Driesse:2024xad} this prescription, while finite, fails to reproduce conservative terms, which are odd in $v$, stemming from the tail-of-tail \cite{Dlapa:2025biy}.

At the 5PM-2SF order, the potential region is by construction not sensitive to the $\iO$ prescription, and evaluating the tail region using Feynman propagators indeed cancels the $\epsilon$-pole coming from the potential region. Yet it leaves the novel singularity at $\gamma=3$ intact that can only be cancelled from the memory region, which in turn should not introduce new $\epsilon$-poles. 
In fact, in the memory region there are only two boundary integrals appearing:
\be
I^{(M)}_{1}=
\raisebox{-0.3cm}{
\begin{tikzpicture}[scale=0.8]
    \coordinate (inA)  at (0.6,0.6);
    \coordinate (outA) at (2.625,0.6);
    \coordinate (inB)  at (0.6,-0.3);
    \coordinate (outB) at (2.625,-0.3);

    \draw[dotted] (inA) -- (outA);
    \draw[dotted] (inB) -- (outB);

    \coordinate (aA) at (0.6,0.6);
    \coordinate (bA) at (1.6125,0.6);
    \coordinate (cA) at (1.95,0.6);
    \coordinate (dA) at (2.625,0.6);
    \coordinate (eA) at (3.3,0.6);
    
    \coordinate (aM) at (0.6,0);
    \coordinate (bM) at (1.275,0);
    \coordinate (cM) at (1.95,0);
    \coordinate (dM) at (2.625,0);
    \coordinate (eM) at (3.3,0);
    
    \coordinate (aMd) at (0.6,-0.1);
    \coordinate (bMd) at (1.275,-0.1);
    \coordinate (cMd) at (1.95,-0.1);
    \coordinate (dMd) at (2.625,-0.1);
    \coordinate (eMd) at (3.3,-0.1);

    \coordinate (aMdd) at (0.6,-0.15);
    \coordinate (bMdd) at (1.275,-0.15);
    \coordinate (cMdd) at (1.95,-0.15);
    \coordinate (dMdd) at (2.625,-0.15);
    \coordinate (eMdd) at (3.3,-0.15);

    \coordinate (aMu) at (0.6,0.1);
    \coordinate (bMu) at (1.275,0.1);
    \coordinate (cMu) at (1.95,0.1);
    \coordinate (dMu) at (2.625,0.1);
    \coordinate (eMu) at (3.3,0.1);
    
    \coordinate (aB) at (0.6,-0.3);
    \coordinate (bB) at (1.6125,-0.3);
    \coordinate (cB) at (1.95,-0.3);
    \coordinate (dB) at (2.625,-0.3);
    \coordinate (eB) at (3.3,-0.3);

    % vertical photons: middle two red
    \draw[photon]     (aA) -- (aB) ;
    \draw[photon,red] (aA) to[out=80, in=100] (bA);
    \draw[photon,red] (bA) to[out=80, in=100] (dA);
    \draw[photon] (dA) -- (dB);
    \draw[photon] (bA) -- (bB);
    
    % top thick segment only between b and c
%    \draw[zUndirected] (bA) -- (cA);

    % vertices
    \foreach \v in {aA,bA,dA,aB,bB,dB}
      \draw[fill] (\v) circle (.08);
  \end{tikzpicture}}\, , \qquad 
I^{(M)}_{2}=
\raisebox{-0.3cm}{
\begin{tikzpicture}[scale=0.8]
    \coordinate (inA)  at (0.6,0.6);
    \coordinate (outA) at (2.625,0.6);
    \coordinate (inB)  at (0.6,-0.3);
    \coordinate (outB) at (2.625,-0.3);

    \draw[dotted] (inA) -- (outA);
    \draw[dotted] (inB) -- (outB);

    \coordinate (aA) at (0.6,0.6);
    \coordinate (bA) at (1.6125,0.6);
    \coordinate (cA) at (1.95,0.6);
    \coordinate (dA) at (2.625,0.6);
    \coordinate (eA) at (3.3,0.6);
    
    \coordinate (aM) at (0.6,0);
    \coordinate (bM) at (1.275,0);
    \coordinate (cM) at (1.95,0);
    \coordinate (dM) at (2.625,0);
    \coordinate (eM) at (3.3,0);
    
    \coordinate (aMd) at (0.6,-0.1);
    \coordinate (bMd) at (1.275,-0.1);
    \coordinate (cMd) at (1.95,-0.1);
    \coordinate (dMd) at (2.625,-0.1);
    \coordinate (eMd) at (3.3,-0.1);

    \coordinate (aMdd) at (0.6,-0.15);
    \coordinate (bMdd) at (1.275,-0.15);
    \coordinate (cMdd) at (1.95,-0.15);
    \coordinate (dMdd) at (2.625,-0.15);
    \coordinate (eMdd) at (3.3,-0.15);

    \coordinate (aMu) at (0.6,0.1);
    \coordinate (bMu) at (1.275,0.1);
    \coordinate (cMu) at (1.95,0.1);
    \coordinate (dMu) at (2.625,0.1);
    \coordinate (eMu) at (3.3,0.1);
    
    \coordinate (aB) at (0.6,-0.3);
    \coordinate (bB) at (1.6125,-0.3);
    \coordinate (cB) at (1.95,-0.3);
    \coordinate (dB) at (2.625,-0.3);
    \coordinate (eB) at (3.3,-0.3);

    % vertical photons: middle two red
    \draw[photon]     (aA) -- (aB) ;
    \draw[photon,red] (aA) to[out=80, in=100]  (bA) ;
    \draw[photon,red] (bA) to[out=80, in=100]  (dA);
    \draw[photon] (dA) -- (dB);
    \draw[photon] (bA) -- (bB);
    \draw[fill,red] (1.12,0.9) circle (.1);
     \draw[fill,red] (2.14,0.9) circle (.1);
    
    % top thick segment only between b and c
%    \draw[zUndirected] (bA) -- (cA);
    % vertices
    \foreach \v in {aA,bA,dA,aB,bB,dB}
      \draw[fill] (\v) circle (.08);
  \end{tikzpicture}}\, ,
\ee
where the red dots indicate squared graviton propagators.
Intriguingly, \emph{imposing} the cancellation
of the $\gamma=3$ singularity along with maintaining the $\epsilon$-pole cancellation
between the potential and tail regions determines their results up to a single 
undetermined coefficient $c_{M}$:
\begin{align}
\label{eq:I12}
\begin{aligned}
I^{(M)}_{1}&=  \frac{1}{ 15 (8 \pi)^4 \epsilon } +\mathcal{O}(\epsilon^0)\,, \,
I^{(M)}_{2}= - \frac{5\, c_{M}}{ 6 (8 \pi)^4  \epsilon^{2}}
 +\mathcal{O}(\epsilon^{-1}) .
\end{aligned}
\end{align}
However, a calculation for the Feynman $\iO$ prescription 
yields an $\epsilon^{-4}$ pole for $I^{(M)}_{2}$ which breaks finiteness of the final
result through a $\eps^{-2}$ pole and a $(\gamma-3)^{-15/2}$ divergence.

Hence, we need to redefine the conservative prescription in such a fashion
that the following three conditions are met: (i) cancellation of dimensional regulator poles,
(ii) cancellation of $(\gamma-3)$-divergence and (iii) time reflection symmetry. 
Such a prescription would provide \eqn{eq:I12} with a given $c_{M}$.
Interestingly, we observe that evaluating $I^{(M)}_{1,2}$ with retarded
propagators pointing towards the middle point 
provides such a prescription --- that we term ``$\gamma$-3’’ --- and leads to the value
$c_{M}=1$, together with a finite impulse, see supplementary material for details.  
Yet we acknowledge that the physical motivation for this prescription is opaque and may not capture all conservative effects. We leave it to future work to elucidate this
important question.

%===========================
\sec{Function space} 
The final function space includes the K3 period 
$\varpi_\text{K3}(x)$, its derivative $\varpi'_\text{K3}(x)$, and up to three times iterated integrals 
\begin{align}\label{IIdef}
  \mathcal I [\varphi_1,...,\varphi_n;x]=\int_1^x \mathrm dx’\,\varphi_1(x')\,\mathcal{I}[\varphi_2,...,\varphi_n;x’] \, ,
\end{align}
where the integration kernels $\varphi_i(x)$ are selected from the set
\begin{align}
	\bigg\{ &\dfrac{1}{x}, \dfrac{1}{1+x^2}, \dfrac{1+x}{x(x-1)}, \dfrac{-1+x}{x(x+1)}, \dfrac{1-x^2}{x+x^2+x^3}, \dfrac{1-x^2}{x-x^2+x^3}, \nonumber \\ 
	&\quad \dfrac{-1+x^2}{x+x^3}, \dfrac{(1+x^2)\,\varpi_\text{K3}(x)}{x}, \dfrac{1}{x \,r(x)\,\varpi_\text{K3}(x)}\bigg\} 
\end{align}
with the square root $r(x) = \sqrt{-1+34x^2-x^4}$. The normalised K3 period is chosen to have an even velocity expansion $(1+x^2) \varpi_\text{K3}(x) =1+\frac{15 v^2}{32}+\frac{351 v^4}{1024}+\mathcal O(v^6)$.
Even though we encounter CY3 varieties in the DEs, the integrals associated with this geometry drop out after IBP reduction of the full integrand in $D$-dimensions and do not contribute to the final result.

\sec{Results}
The scattering angle $\theta_{\text{cons}}$ follows from the impulse using $|\Delta p^\mu_{i,\rm cons}|=2p_\infty\sin(\theta_{\text{cons}}/2)$.
Here $p_\infty=m_1 m_2 \sqrt{\gamma^2-1}/E$, the total (conserved) energy is $E=M\sqrt{1+2\nu(\gamma-1)}$ and the total mass is $M=m_1+m_2$, with $\nu=m_1m_2/M^2$ the symmetric mass ratio.
The scattering angle is given in PM expansion as 
\begin{align}
  \label{thetadef}
    \theta_{\text{cons}}=\frac{E}{M} \sum_{m\geq1}\sum_{s=0}^{\lfloor\frac{m-1}{2}\rfloor}
    \bigg(
      \frac{GM}{|b|}
      \bigg)^m
      \nu^s
    \theta^{(m,s)}_{\text{cons}}(\gamma)\,,
  \end{align}
  where $s$ counts the SF order.
Our main result is the 5PM-2SF contribution 
\begin{align}
  \theta^{(5,2)}_{\text{cons}}
  =
  \sum_{k=1}^{36}
  c_k(\gamma) f_k(\gamma)\,,
\end{align}
where $f_k(\gamma)$ are the 36 linear combinations of iterated integrals discussed in the previous section and $c_k(\gamma)$ are polynomials in $\gamma$ and $\sqrt{\gamma^2-1}=\gamma v$, see the tables in the 
supplementary material. We present all our analytical results in the accompanying \Zenodo submission, {thereby tagging the regions and also providing explicit
PN expansions up to $v^{500}$ for numerical evaluations.}

\begin{figure}[t] 
	\includegraphics[width=0.9\hsize]{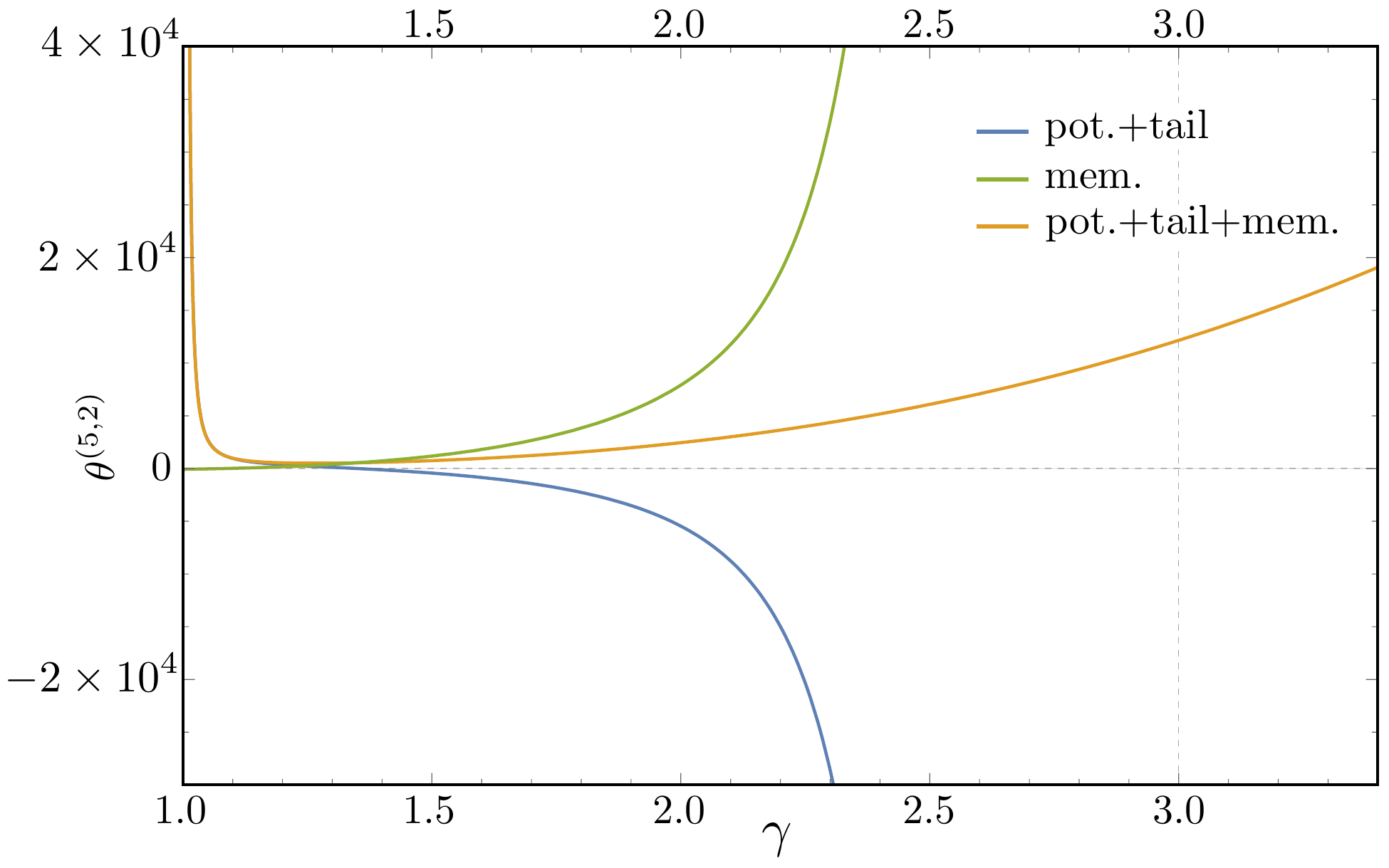}
	 \caption{The 5PM-2SF contribution to the scattering angle, $\theta^{(5,2)}_{\text{cons}}(\gamma)$: 
	 Potential and memory contributions both diverge for $\gamma \rightarrow 3$. These divergences cancel for the full result if one uses \eqn{eq:I12} irrespective of
	 the value of $c_{M}$, which is set to 1 for this plot.
	 }
	  \label{fig-angle}
	  \end{figure}
%===========================

\sec{Checks}
Our 5PM angle agrees in the low-velocity $v=\sqrt{\gamma^2-1}/\gamma\to0$ limit with the scattering 
angle up to 4PN order~\cite{Bini:2020hmy,Blumlein:2020pog,Bini:2021gat,Bini:2022enm,Bini:2025vuk}.
Up to 5PN order with the memory contribution using \eqref{eq:I12}  in square brackets we find
\begin{align}
%\begin{aligned}
  &\theta^{(5,2)}_{\text{cons}}
  = \dfrac{2}{5 v^6}+\dfrac{53}{5 v^4}+ \left (\dfrac{45341}{360}-\dfrac{41 \pi^2}{12}\right )\dfrac{1}{v^{2}}  + \dfrac{46629199}{15120}\nonumber \\ & 
   + \left [ \frac{64 \,c_{M}}{5}-\frac{11456}{135} \right ]-\dfrac{221597 \pi ^2}{720} + \dfrac{2816 \log[2v]}{45} 
+\mathcal{O}(v^2)
 \, .
% \end{aligned}
  \end{align}
Our $\cO(v^{0})$ result confirms a conjecture on the $\pi^{2}$ 5PN contributions
being purely potential~\cite{Blumlein:2020pyo,Bini:2025vuk}, while our rational 
tail contribution at this order  agrees with the 5PN tail contribution of
\Rcite{PortoPrivate} based on \Rcite{Porto:2024cwd}.
Restricting to the potential region we agree with the very recent low-velocity expanded result for the scattering angle of \Rcite{Bern:2025wyd}.
Finally, the discontinuity of the scattering angle is related to the radiated energy at one order lower in the PM expansion:
\begin{align}\label{eq:discontinuity}
    \frac{\theta_{\rm cons}(-\gamma_{-}\!+\!\i\eps)-\theta_{\rm cons}(-\gamma_{-} \! -\! \i\eps)}{2i\pi}
    =
      G E
      \frac{\partial E_{\rm rad}|_{{\rm odd-in-}v}}{
        \partial L
      }
      %\bigg|_{{\rm odd-in-}v}\!\!\!\!\!\!
  \end{align}
where $\gamma_-=\gamma-1$,  and the total angular
momentum is $L=\pin |b|$~\cite{Cho:2021arx,Dlapa:2021vgp,Jakobsen:2023hig,Jakobsen:2023pvx,Bini:2017wfr}.

%===========================
\sec{Conclusions} 
In this Letter, we have {computed conservative contributions
to the momentum impulse and scattering angle}
in the non-spinning gravitational two-body problem
at {5PM ($G^{5}$) and 2SF} order using WQFT.
The complexity of this extraordinarily challenging computation
{was an order of magnitude larger than the 1SF conservative computation~\cite{Driesse:2024xad},}
and could only be tackled by greatly innovating our integration strategies.
The emerging function space is richer than {that of the 1SF case
(which involved only polylogarithms up to weight 3)
due to the emergence of a K3 period.}

{A crucial discovery of our work was a spurious divergence
existing at $v/c=\sqrt{8}/3$, i.e.~$\gamma=3$,
which} needs to cancel between the potential and memory regions.
Yet the present prescription for extracting conservative data
using Feynman propagators,
{previously adopted at 5PM-1SF order~\cite{Driesse:2024xad},}
\emph{fails} to provide a finite result.
Enforcing {cancellation} leads to a novel $\gamma$-3 prescription that we propose.
This situation calls for a {better} definition of the conservative sector in BH scattering, perhaps relying on the operator $\hat{N}=-i\log\hat{S}$~\cite{Damgaard:2021ipf, Damgaard:2023ttc, Kim:2024svw, Alessio:2025flu, Kim:2025gis, Haddad:2025cmw, Brandhuber:2025igz}. Clearly, we should next turn to the fully dissipative 5PM-2SF computation, as here there is an unambiguous understanding of the worldline formalism using retarded propagators throughout \Rcites{Jakobsen:2022psy, Kalin:2022hph}.

\sec{Acknowledgements}
We thank Thibault Damour, Riccardo Gonzo, Albrecht Klemm, Rafael Porto and Lorenzo Tancredi for insightful discussions.
We also thank Rafael Porto for communicating unpublished results on the 5PN tail contributions as well as 
Agostino Patella for advice on HPC. 
This work was funded by the Deutsche Forschungsgemeinschaft
(DFG, German Research Foundation)
Projektnummer 417533893/GRK2575 ``Rethinking Quantum Field Theory'' (GUJ, GM, BS, JP, JU) 
and 508889767/FOR5372/1 ``Modern Foundations of Scattering Amplitudes'' (JP),
by The Royal Society under grant URF\textbackslash R1\textbackslash 231578,
``Gravitational Waves from Worldline Quantum Field Theory'' (GM),
and by the European Union through the 
European Research Council under grant ERC Advanced Grant 101097219 (GraWFTy) (MD, JP, JU),
the ERC Synergy Grant 101167287 (MaScAmp) (CN)
 and ERC Starting Grant 949279 (HighPHun) (CN).
Views and opinions expressed are, however, those of the authors only and do not necessarily reflect those of the European Union or European Research Council Executive Agency. Neither the European Union nor the granting authority can be held responsible for them.
The authors gratefully acknowledge the computing time made available to them on the 
high-performance computer Lise at the NHR Center Zuse-Institut Berlin (ZIB). This center is jointly supported by the Federal Ministry of Education and Research and the state governments participating in the National High-Performance Computing (NHR) joint funding program 
(http://www.nhr-verein.de/en/our-partners).

\clearpage

\appendix
\begin{widetext}
\section*{Supplementary Material}

%===========================
\sec{Integral families}
The first three categories of integral families {(P, NP1, NP2)} share a common schematic form:
\begin{align}\label{integralSchematic}
  {\cal I}_{\{n\}}^{\{\sigma\}}
  =
  \int_{\ell_1\ell_2\ell_3\ell_4}\!\!\!\!\!\!\!\!
  \frac{
    \prod_{i=1}^{2}\dd^{(\bar{n}_i\!-\!1)}(\ell_i\cdot v_1)
    %\dd^{(\bar{n}_2\!-\!1)}(\ell_2\cdot v_1)
    \prod_{j=3}^{4}\dd^{(\bar{n}_j\!-\!1)}(\ell_j\cdot v_2)
    %\dd^{(\bar{n}_4\!-\!1)}(\ell_4\cdot v_2)
    }{
    \prod_{i}D_{i}^{n_{i}}(\sigma_{i})
    \prod_{I<J}D_{IJ}^{n_{IJ}}
    }\,,
\end{align}
where $\{\sigma\}$ and $\{n\}$ are collections of $\i 0^+$ signs and integer powers of propagators, respectively. We also write $\int_{\ell}= \int\!\!\frac{\mathrm d^{D}\ell}{(2\pi)^{D}}$ 
as well as $\dd^{n}(x)=2\pi (\partial/\partial x)^{n}\delta(x)$.
There are six possible worldline propagators:
\begin{align}
\begin{aligned}
  D_1(\sigma_1)&=\ell_1\cdot v_2+\sigma_1 \i 0^+\,, \qquad\qquad
  D_2(\sigma_2)=\ell_2\cdot v_2+\sigma_2 \i 0^+\,,\qquad \qquad
  D_3(\sigma_3)=(\ell_1-\ell_2)\cdot v_2+\sigma_3\i 0^+\,,\\
  D_4(\sigma_4)&=\ell_3\cdot v_1+\sigma_4 \i 0^+\,, \qquad\qquad
  D_5(\sigma_5)=\ell_4\cdot v_1+\sigma_5 \i 0^+\,, \qquad\qquad
  D_6(\sigma_6)=(\ell_3-\ell_4)\cdot v_1+\sigma_6 \i 0^+\,.
\end{aligned}
\end{align}
The massless bulk propagators $D_{IJ}$ with $I=(0,i,q)$ are (ignoring the Feynman $\i 0^{+}$ prescription) 
\begin{align}\label{DIJs}
  D_{ij}&=(\ell_i-\ell_j)^2
  \,,&
  D_{qi}&=(\ell_i+q)^2
  \,,&
  D_{0i}&=\ell_i^2
  \,.
\end{align}
Inclusion of six worldline propagators $D_i$ makes this an over-complete basis, as those propagators satisfy linear identities.
These imply partial fractions:
\begin{align}\label{partialFractions}
\begin{aligned}
  \frac{1}{D_1(\sigma_1)D_2(\sigma_2)}=\frac{1}{D_2(\sigma_2)D_3(\sigma_3)}-\frac{1}{D_1(\sigma_1)D_3(\sigma_3)}\,,\qquad
  \frac{1}{D_4(\sigma_4)D_5(\sigma_5)}=\frac{1}{D_5(\sigma_5)D_6(\sigma_6)}-\frac{1}{D_4(\sigma_4)D_6(\sigma_6)}\,,
\end{aligned}
\end{align}
which hold regardless of the $\i0^+$ prescriptions ($\sigma_i$).

In order to perform IBP reductions, we identify three subsets of the
worldline propagators corresponding to the {P, NP1, and NP2 integral families:
\begin{subequations}\label{extendedFamilies}
\begin{align}
  {\rm P}&: \qquad \{D_1\,,D_2\,,D_4\,,D_5\}\,, \\
  {\rm NP1}&: \qquad  \{D_1\,,D_3\,,D_4\,,D_5\}\,, \\
  {\rm NP2}&: \qquad  \{D_1\,,D_3\,,D_4\,,D_6\}\,.
\end{align}
\end{subequations}}
By repeated use of the partial fraction identities~\eqref{partialFractions} (and symmetries), any integral whose worldline propagators do not fall into one of the subsets in eq.~\eqref{extendedFamilies} may be written as a linear combination of those that do.
In this context, the planar ``P" family is precisely analogous to the Mondrian 1SF family
of Eqn.~(5) in \cite{Driesse:2024xad}.
Inclusion of the extra {NP1 and NP2} families allows for crossings of the bulk gravitons produced by the inclusions of worldlines, which --- unlike at 1SF --- cannot be entirely eliminated by partial fractions.
Symmetries of these three families amount to shifts of the loop momenta $\ell_{i}$ by the momentum transfer $q$ as well as $S_{2}$ permutations of $\ell_{i}$ and $v_{i}$.

{The other integral family is NPM,
which is nonplanar and associated with nonlinear memory.
The NPM} family has the same schematic form as in eq.~\eqref{integralSchematic},
but with the following worldline propagators:
{\begin{align}
\begin{aligned}
  D^{\rm NPM}_1&=\ell_1\cdot v_2+\sigma_1 \i 0^+\,, &
  D^{\rm NPM}_2&=(\ell_1-\ell_2)\cdot v_2+\sigma_2 \i 0^+\,,\\
  D^{\rm NPM}_3&=\ell_3\cdot v_1+\sigma_3 \i 0^+\,, &
  D^{\rm NPM}_4&=\ell_4\cdot v_1+\sigma_4 \i 0^+\,.
\end{aligned}
\end{align}}
The bulk propagators now have a different form:
{\begin{align}
\begin{aligned}
  D^{\rm NPM}_5&=(\ell_1-\ell_2-\ell_3+\ell_4)^2\,,	&& &&\\
  D^{\rm NPM}_6&=(\ell_1-\ell_3)^2\,, \qquad\qquad
  &&D^{\rm NPM}_7=(\ell_2-\ell_4)^2\,,				&& &&\\
  D^{\rm NPM}_8&=(\ell_1-\ell_2)^2\,, \qquad\qquad
  &&D^{\rm NPM}_9=(\ell_3-\ell_4)^2\,,				&&\\
  D^{\rm NPM}_{10}&=\ell_1^2\,, \quad
  D^{\rm NPM}_{11}=\ell_2^2\,, \quad
  &&D^{\rm NPM}_{12}=\ell_3^2\,, \quad
  &&D^{\rm NPM}_{13}=\ell_4^2\,,\\
  D^{\rm NPM}_{14}&=(\ell_1+q)^2\,, \qquad\qquad
  &&D^{\rm NPM}_{15}=(\ell_3+q)^2\,,				&&\\
  D^{\rm NPM}_{16}&=\ell_2\cdot\ell_3\,, \qquad\qquad
  &&D^{\rm NPM}_{17}=\ell_2\cdot q\,, \qquad
  &&D^{\rm NPM}_{18}=\ell_4\cdot q\,.
  \label{DNPs}
\end{aligned}
\end{align}}
The three terms {$D_{16}^{\rm NPM}$, $D_{17}^{\rm NPM}$, $D_{18}^{\rm NPM}$} are not ``propagators'' per se --- as the corresponding top-sectors are not populated, we choose simple dot products to improve IBP performance.
The graph (k) of Fig.~\ref{fig:diagrams} displays a richer $S_{3}$ permutation symmetry: permutations of the three active graviton legs meeting in the central three-point vertex leave it invariant. 
We employ all these permutation symmetries to maximally simplify the integrand as well as the IBP reduction.

\sec{Magic relations in IBP reduction} 
To bring the system of DEs into $\eps$-form, we cast the master-integral basis into a denominator-factorised form \cite{Usovitsch:2020jrk, Smirnov:2020quc}, first. An efficient way to construct such a factorised basis is to set all subsectors to zero. When studying individual sectors, however, we found that the number of master integrals was off by one. To recover the correct count, we extended the reduction to include higher sectors, which generate the required magic relations \cite{Maierhofer:2018gpa}. This issue only became apparent at that stage. All other reductions were performed with full sector dependence.

\sec{Boundary Integration}
In a few cases, we apply a ``Feynman parameter integration through differential equations" strategy (FP) of \cite{Hidding:2022ycg} to compute the missing integrals numerically. From high-precision numerical data, we utilise the PSLQ algorithm \cite{Ferguson1999PSLQ} to reconstruct the analytic results. The core ingredient of the FP strategy is to introduce an auxiliary parameter $y$ into the parameter free boundary integrals by the Feynman parameterising: $A^{-a}B^{-b}=\frac{\Gamma[a+b])}{\Gamma[a] \Gamma[b]} \int_0^\infty dy\,y^{-1+b} \,(A+B y)^{-a-b}$. The two necessary ingredients for the FP strategy are the analytic calculation of the system of DEs with respect to the auxiliary parameter $y$ and the calculation of the boundary conditions analytically in the limit where the auxiliary parameter $y$ vanishes. With both ingredients equipped, we use \texttt{AmpRed} \cite{Chen:2024xwt, Chen:2019mqc, Chen:2019fzm, Chen:2020wsh}, which allows us to use a user-defined system of DEs and boundary conditions, to compute the definite integral in the auxiliary parameter $y$ in the interval $[0,\infty]$ numerically, which yields the original integral in return. 

Furthermore, we observe that the system of DEs with respect to the auxiliary parameter can be put into $\epsilon$-form. Together with the analytic boundary conditions at $y=0$, we have computed the definite integral analytically with \texttt{PolyLogTools}~\cite{Duhr:2019tlz}, which resolves the solution in terms of the same result as what we have reconstructed with the help of FP + \texttt{AmpRed} + PSLQ.

\sec{Memory boundary integrals}
For the memory graph $I_1^{(M)}$, we find the following momentum space expression:
\begin{align}
  I_1^{(M)}
  =
  \int_{\ell_1\ell_2\ell_3\ell_4}
  \frac{
    \dd(\ell_1\cdot v_1)
    \dd(\ell_2\cdot v_1)
    \dd(\ell_3\cdot v_2)
    \dd(\ell_4\cdot v_2)
  }{D_{q3}D_{34}D_{4}D_{31}D_{24}}
  \ .
\end{align}
In order to use the $\gamma$-3 prescription, one must use a retarded prescription for the two active denominators in this integrand: $D_{31}$ and $D_{24}$.
We will evaluate the integral in a perturbative series in $\epsilon$ using a frequency domain method.
This method does not work well for $I_2^{(M)}$ and instead we re-express it in terms of $I_1^{(M)}$ and a third memory integral $I_3^{(M)}$ which has the same denominator structure as $I_2^{(M)}$ but is dressed with scalar products. Its expression is,
\begin{align}
  I_{3}^{(M)}
  =
  -\int_{\ell_1\ell_2\ell_3\ell_4}
  \frac{D_4^3 D_5^3
    \dd(\ell_1\cdot v_1)
    \dd(\ell_2\cdot v_1)
    \dd(\ell_3\cdot v_2)
    \dd(\ell_4\cdot v_2)
  }{D_{q3}D_{34}D_{4}D_{31}^2D_{24}^2}
  \ ,
\end{align}
where, again, retarded propagators should be used for $D_{31}$ and $D_{24}$ for the $\gamma$-3 prescription.
The relevant IBP relation involving these three integrals reads
\begin{align}\label{eq:IBP}
  I_2^{(M)}=&\frac{28 (5 \epsilon -2) (14 \epsilon -9) \left(1088 \epsilon ^4-898 \epsilon ^3+275 \epsilon ^2-38 \epsilon +2\right) I_3^{(M)}}{3 \epsilon ^3 \left(96 \epsilon ^3-116 \epsilon ^2+43 \epsilon -5\right)}\nn \\
  +&\frac{(1-2 \epsilon )^2 \left(121024 \epsilon ^6-237340 \epsilon ^5+181552 \epsilon ^4-70363 \epsilon ^3+14767 \epsilon ^2-1612 \epsilon +72\right) I_1^{(M)}}{6 \epsilon ^3 \left(96 \epsilon ^3-116 \epsilon ^2+43 \epsilon -5\right)} \, .
\end{align}

We evaluate the integrals using the following two ingredients:
\begin{equation}
  \raisebox{-0.3cm}{
\begin{tikzpicture}[scale=0.8]
    \coordinate (inA)  at (1.2,0.6);
    \coordinate (outA) at (2.025,0.6);
    \coordinate (inB)  at (1.2,-0.3);
    \coordinate (outB) at (2.025,-0.3);

    \coordinate (aA) at (0.6,0.6);
    \coordinate (bA) at (1.6125,0.6);
    \coordinate (cA) at (1.95,0.6);
    \coordinate (dA) at (2.625,0.6);
    \coordinate (eA) at (3.3,0.6);
    
    \coordinate (aM) at (0.6,0);
    \coordinate (bM) at (1.275,0);
    \coordinate (cM) at (1.95,0);
    \coordinate (dM) at (2.625,0);
    \coordinate (eM) at (3.3,0);
    
    \coordinate (aMd) at (0.6,-0.1);
    \coordinate (bMd) at (1.275,-0.1);
    \coordinate (cMd) at (1.95,-0.1);
    \coordinate (dMd) at (2.625,-0.1);
    \coordinate (eMd) at (3.3,-0.1);

    \coordinate (aMdd) at (0.6,-0.15);
    \coordinate (bMdd) at (1.275,-0.15);
    \coordinate (cMdd) at (1.95,-0.15);
    \coordinate (dMdd) at (2.625,-0.15);
    \coordinate (eMdd) at (3.3,-0.15);

    \coordinate (aMu) at (0.6,0.1);
    \coordinate (bMu) at (1.275,0.1);
    \coordinate (cMu) at (1.95,0.1);
    \coordinate (dMu) at (2.625,0.1);
    \coordinate (eMu) at (3.3,0.1);
    
    \coordinate (aB) at (0.6,-0.3);
    \coordinate (bB) at (1.6125,-0.3);
    \coordinate (cB) at (1.95,-0.3);
    \coordinate (dB) at (2.625,-0.3);
    \coordinate (eB) at (3.3,-0.3);

    % vertical photons: middle two red
    % \draw[photon]     (aA) -- (aB) ;
    % \draw[photon,red] (aA) to[out=80, in=100] (bA);
    % \draw[photon,red] (bA) to[out=80, in=100] (dA);
    % \draw[photon] (dA) -- (dB);
    \draw[dotted] (inA) -- (bA);
    \draw[dotted] (inB) -- (outB);

    \draw[zParticleF] (bA) -- (outA) ;

    \draw[photon] (bA) -- (bB);

    \node[below] at (outA) {$ \omega $} ;
    
    % top thick segment only between b and c
%    \draw[zUndirected] (bA) -- (cA);

    % vertices
    \foreach \v in {bA,bB}
      \draw[fill] (\v) circle (.08);
  \end{tikzpicture}}
  =
  -
  (2\pi)^{\epsilon-1}
  \bigg(
    \frac{|b|}{|\omega|}
  \bigg)^{\epsilon}
  K_{\epsilon}(|b||\omega|)
  \ ,
  \qquad
    \raisebox{-0.3cm}{
\begin{tikzpicture}[scale=0.8]
    \coordinate (inA)  at (.17,0.6);
    \coordinate (outA) at (2.025,0.6);
    \coordinate (inB)  at (1.2,-0.3);
    \coordinate (outB) at (2.025,-0.3);

    \coordinate (aA) at (0.6,0.6);
    \coordinate (abA)at (1.1,.9) ;
    \coordinate (bA) at (1.6125,0.6);
    \coordinate (cA) at (1.95,0.6);
    \coordinate (dA) at (2.625,0.6);
    \coordinate (eA) at (3.3,0.6);
    
    \coordinate (aM) at (0.6,0);
    \coordinate (bM) at (1.275,0);
    \coordinate (cM) at (1.95,0);
    \coordinate (dM) at (2.625,0);
    \coordinate (eM) at (3.3,0);
    
    \coordinate (aMd) at (0.6,-0.1);
    \coordinate (bMd) at (1.275,-0.1);
    \coordinate (cMd) at (1.95,-0.1);
    \coordinate (dMd) at (2.625,-0.1);
    \coordinate (eMd) at (3.3,-0.1);

    \coordinate (aMdd) at (0.6,-0.15);
    \coordinate (bMdd) at (1.275,-0.15);
    \coordinate (cMdd) at (1.95,-0.15);
    \coordinate (dMdd) at (2.625,-0.15);
    \coordinate (eMdd) at (3.3,-0.15);

    \coordinate (aMu) at (0.6,0.1);
    \coordinate (bMu) at (1.275,0.1);
    \coordinate (cMu) at (1.95,0.1);
    \coordinate (dMu) at (2.625,0.1);
    \coordinate (eMu) at (3.3,0.1);
    
    \coordinate (aB) at (0.6,-0.3);
    \coordinate (bB) at (1.6125,-0.3);
    \coordinate (cB) at (1.95,-0.3);
    \coordinate (dB) at (2.625,-0.3);
    \coordinate (eB) at (3.3,-0.3);

    % vertical photons: middle two red
    % \draw[photon]     (aA) -- (aB) ;
    \draw[photon,red] (aA) to[out=80, in=100] (bA);
    % \draw[photon,red] (bA) to[out=80, in=100] (dA);
    % \draw[photon] (dA) -- (dB);
    \draw[dotted] (aA) -- (bA);
%    \draw[dotted] (inB) -- (outB);

    \draw[zParticleF] (bA) -- (outA) ;
    \draw[zParticleF] (inA) -- (aA) ;

    % \draw[photon] (bA) -- (bB);

    \node[below] at (outA) {$ \omega $} ;
    \node[below] at (inA) {$ \omega $} ;
    \node[above] at (abA) {$ a $} ;
    
    % top thick segment only between b and c
%    \draw[zUndirected] (bA) -- (cA);

    % vertices
    \foreach \v in {bA,aA}
      \draw[fill] (\v) circle (.08);
  \end{tikzpicture}}
  =
  (-1)^a \frac{
    \Gamma(\frac{2a-3+2\epsilon}{2})
  }{
    \Gamma(a)
    \sqrt{4\pi}^{3-2\epsilon}
  }
  (0^+
  -
  \i\omega)^{3-2a-2\epsilon}
  \ .
\end{equation}
In the first diagram, we are in $b$-space rather than $q$-space and $K_\epsilon(z)$ is the modified Bessel function of the second kind.
Further, we have set $\gamma=\sqrt{2}$ in order to simplify the known gamma dependence --- overall powers of $\sqrt{\gamma^2-1}$ --- of boundary integrals (and likewise in the IBP relation Eq.~\eqref{eq:IBP}).

In this way, we arrive at the following two expressions for $I_1^{(M)}$ and $I_3^{(M)}$:
\begin{align}
  I_1^{(M)}
  &=
  - 2^{7\epsilon-1}
  \frac{
    \Gamma^2(\frac{2\epsilon-1}{2})\Gamma(4\epsilon-1)
    }{
    \Gamma(2-5\epsilon)(4\pi)^{5-4\epsilon}
  }
  \int_{\omega_1, \omega_2}
  (0^+-\i \omega_1)^{1-2\epsilon}
  (0^+-\i \omega_2)^{1-2\epsilon}
  \prod_{i=1}^3
  |\omega_i|^\epsilon K_\epsilon(|\omega_i|)
%  \mathcal K(\omega_1,\omega_2) 
%  |\omega_1\omega_2(\omega_1+\omega_2)|^{-\epsilon}
%  K_{\epsilon}(|\omega_1|)
%  K_{\epsilon}(|\omega_2|)
%  K_{\epsilon}(|\omega_1+\omega_2|)
  \ ,
  \\
  I_3^{(M)}
  &=
  2^{-3+7\epsilon}
  \frac{
    \Gamma^2(\frac{2\epsilon+1}{2})\Gamma(4\epsilon-2)
    }{
    \Gamma(3-5\epsilon)(4\pi)^{5-4\epsilon}
  }
  \int_{\omega_1, \omega_2}
  \omega_1^3\omega_2^3
  (0^+-\i \omega_1)^{-1-2\epsilon}
  (0^+-\i \omega_2)^{-1-2\epsilon}
    \prod_{i=1}^3
  |\omega_i|^\epsilon K_\epsilon(|\omega_i|)
  \ ,
 \end{align}
with $\int_\omega = \int_{-\infty}^\infty\frac{\mathrm d\omega}{2\pi}$ and $\omega_3=\omega_1+\omega_2$. 
As it turns out, the frequency integrations in both of these expressions converge in the $\epsilon\to0$ limit.
For this reason, we Taylor expand the integrand in $\epsilon$ before integration.
We require both integrals at sub-sub-leading order in $\epsilon$, and we thus need to expand the integrand two orders beyond the leading order.
In particular, we use:
\begin{align}
  (0^+-\i\omega)^{-2\epsilon}
  =
  e^{-2\epsilon \log(0^+-\i\omega)}
  =
  1
  -2\epsilon
  \log(0^+-\i\omega)
  +
  2\epsilon^2 \log^2(0^+-\i\omega)
  +
  \dots
\end{align}
where
\begin{align}
  \log(0^+-\i\omega)
  =
  \log|\omega|-\frac{\i\pi}2 \text{sign}( \omega )
  \ .
\end{align}
We generally evaluate the frequency integral numerically.

\sec{The $\gamma$-3 prescription for the memory region}
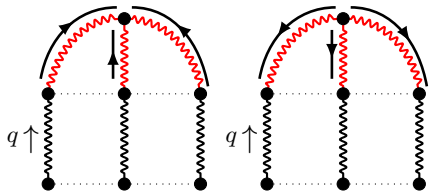
\begin{figure}[t!]
  \begin{tikzpicture}[scale=1]
   \coordinate (inA)  at (0.6,0.6);
   \coordinate (outA) at (2.625,0.6);
   \coordinate (inB)  at (0.6,-0.6);
   \coordinate (outB) at (2.625,-0.6);

   \draw[dotted] (inA) -- (outA);
   \draw[dotted] (inB) -- (outB);

   \coordinate (aA) at (0.6,0.6);
   \coordinate (bA) at (1.6125,0.6);
   \coordinate (bAA) at (1.6125,1.6);
   \coordinate (cA) at (1.95,0.6);
   \coordinate (dA) at (2.625,0.6);
   \coordinate (eA) at (3.3,0.6);
   
   \coordinate (aM) at (0.6,0);
   \coordinate (bM) at (1.275,0);
   \coordinate (cM) at (1.95,0);
   \coordinate (dM) at (2.625,0);
   \coordinate (eM) at (3.3,0);
   
   \coordinate (aMd) at (0.6,-0.1);
   \coordinate (bMd) at (1.275,-0.1);
   \coordinate (cMd) at (1.95,-0.1);
   \coordinate (dMd) at (2.625,-0.1);
   \coordinate (eMd) at (3.3,-0.1);

   \coordinate (aMdd) at (0.6,-0.15);
   \coordinate (bMdd) at (1.275,-0.15);
   \coordinate (cMdd) at (1.95,-0.15);
   \coordinate (dMdd) at (2.625,-0.15);
   \coordinate (eMdd) at (3.3,-0.15);

   \coordinate (aMu) at (0.6,0.1);
   \coordinate (bMu) at (1.275,0.1);
   \coordinate (cMu) at (1.95,0.1);
   \coordinate (dMu) at (2.625,0.1);
   \coordinate (eMu) at (3.3,0.1);
   
   \coordinate (aB) at (0.6,-0.6);
   \coordinate (bB) at (1.6125,-0.6);
   \coordinate (cB) at (1.95,-0.6);
   \coordinate (dB) at (2.625,-0.6);
   \coordinate (eB) at (3.3,-0.6);

   % vertical photons: middle two red
   \draw[photon]     (aA) -- (aB) node [midway,left] {$q\,\,\,\,\,$} node [midway,left] {\mbox{\large$\uparrow$}};
   \draw[photon ,red] (aA) to[out=80, in=-180] (bAA);
   \draw[zParticle] (0.5,0.8) to[out=80, in=-180] (1.5125,1.75);
   \draw[photon ,red] (bAA) to[out=0, in=100] (dA);
   \draw[zParticle] (2.725,0.7) to[out=100, in=-0] (1.7125,1.75);
   \draw[zParticle] (1.4625,0.8) to (1.4625,1.45);
   \draw[photon ,red] (bA) to (bAA);
   \draw[photon] (dA) -- (dB);
   \draw[photon] (bA) -- (bB);
   
   % top thick segment only between b and c
%    \draw[zUndirected] (bA) -- (cA);

   % vertices
   \foreach \v in {aA,bA,dA,aB,bB,dB,bAA}
     \draw[fill] (\v) circle (.08);
 \end{tikzpicture} 
 \begin{tikzpicture}[scale=1]
  \coordinate (inA)  at (0.6,0.6);
  \coordinate (outA) at (2.625,0.6);
  \coordinate (inB)  at (0.6,-0.6);
  \coordinate (outB) at (2.625,-0.6);

  \draw[dotted] (inA) -- (outA);
  \draw[dotted] (inB) -- (outB);

  \coordinate (aA) at (0.6,0.6);
  \coordinate (bA) at (1.6125,0.6);
  \coordinate (bAA) at (1.6125,1.6);
  \coordinate (cA) at (1.95,0.6);
  \coordinate (dA) at (2.625,0.6);
  \coordinate (eA) at (3.3,0.6);
  
  \coordinate (aM) at (0.6,0);
  \coordinate (bM) at (1.275,0);
  \coordinate (cM) at (1.95,0);
  \coordinate (dM) at (2.625,0);
  \coordinate (eM) at (3.3,0);
  
  \coordinate (aMd) at (0.6,-0.1);
  \coordinate (bMd) at (1.275,-0.1);
  \coordinate (cMd) at (1.95,-0.1);
  \coordinate (dMd) at (2.625,-0.1);
  \coordinate (eMd) at (3.3,-0.1);

  \coordinate (aMdd) at (0.6,-0.15);
  \coordinate (bMdd) at (1.275,-0.15);
  \coordinate (cMdd) at (1.95,-0.15);
  \coordinate (dMdd) at (2.625,-0.15);
  \coordinate (eMdd) at (3.3,-0.15);

  \coordinate (aMu) at (0.6,0.1);
  \coordinate (bMu) at (1.275,0.1);
  \coordinate (cMu) at (1.95,0.1);
  \coordinate (dMu) at (2.625,0.1);
  \coordinate (eMu) at (3.3,0.1);
  
  \coordinate (aB) at (0.6,-0.6);
  \coordinate (bB) at (1.6125,-0.6);
  \coordinate (cB) at (1.95,-0.6);
  \coordinate (dB) at (2.625,-0.6);
  \coordinate (eB) at (3.3,-0.6);

  % vertical photons: middle two red
  \draw[photon]     (aA) -- (aB) node [midway,left] {$q\,\,\,\,\,$} node [midway,left] {\mbox{\large$\uparrow$}};
  \draw[photon ,red] (aA) to[out=80, in=-180] (bAA);
  \draw[zParticle] (1.5125,1.75) to[in=80, out=-180] (0.5,0.8);
  \draw[photon ,red] (bAA) to[out=0, in=100] (dA);
  \draw[zParticle] (1.7125,1.75) to[in=100, out=-0]  (2.725,0.7) ;
  \draw[zParticle] (1.4625,1.45) to (1.4625,0.8);
  \draw[photon ,red] (bA) to (bAA);
  \draw[photon] (dA) -- (dB);
  \draw[photon] (bA) -- (bB);
  
  \foreach \v in {aA,bA,dA,aB,bB,dB,bAA}
    \draw[fill] (\v) circle (.08);
\end{tikzpicture}
\caption{
  The $\gamma$-$3$ prescription is given as the average of the two shown causality routings.
  In the first diagram retarded propagators meet in the symmetric point of the three-graviton interaction, and in the second, they are taken as advanced propagators.
  }
\label{fig:g3pres}
\end{figure}
A conservative prescription for the evaluation of the memory boundary conditions leading to physically sensible results must satisfy the three requirements listed in the main text below Eq.~\eqref{eq:I12}.
This is the case for the $\gamma$-3 prescription.
The idea of this prescription is most easily stated for another choice of memory boundary integrals shown in Fig.~\ref{fig:g3pres}, which keep all three radiative propagators.
Upon IBP reduction, one may eliminate (``pinch'') one of the radiative gravitons (e.g., the middle one), and in this way they are expressed in terms of the two memory graphs $I^{(M)}_1$ and $I^{(M)}_2$.
The $\gamma$-3 prescription uses retarded propagators for the memory graphs and averages over two situations: All causality points towards the three-graviton interaction or all causality points away from it.
Crucially, this prescription satisfies equivalent symmetries to a Feynman prescription, including the $S_3$ symmetry of permuting the three subdiagrams joined by the three-graviton interaction.

We note that a region-by-region prescription for picking out conservative effects is currently also necessary at lower PM orders.
Generally, for example, contributions from the R region are ignored, although the real part of using the Feynman prescription gives a non-zero contribution to the scattering angle.
Further, at the 5PM-1SF order, the real part of Feynman also did not produce the expected RRR conservative contributions to the scattering angle.
Interestingly, we note that at the 1SF order, the established real-part-of-Feynman prescription gives equivalent results to an average of using \textit{only} retarded and \textit{only} advanced propagators, analogous to the $\gamma$-3 prescription for the memory.

It is also interesting to note that the self-force approach currently only gives a clear-cut notion of conservative dynamics at the 1SF order, and one may, therefore, not yet seek guidance there.
Further, the appearances of divergences in the Lorentz factor $\gamma$ pose an interesting challenge to the post-Newtonian approach: How can one make sure to avoid such poles when expanding around $\gamma\approx1$?
This is clearly not the case for an angle defined only from potential and tail effects.

\end{widetext}

%\clearpage
\begin{table*}[h!]
	\setlength{\tabcolsep}{1pt} % Default value: 6pt
	\renewcommand{\arraystretch}{3}
	\begin{tabular}{|c|}
		\hline
		\scalebox{0.74}{\tabeq{23cm}{ \\[-0.3cm]
%\hspace{4cm} r(x)=\sqrt{-1+34x^{2}-x^{4}}
%
f^{(E)}_{1}[x] &= 1 \, , \qquad
f^{(E)}_{2}[x] = \pi^{2} \, , \qquad
f^{(E)}_{3}[x] = \II\!\left[\frac{1}{x},\frac{1}{x};x\right] \, , \qquad
f^{(E)}_{4}[x] = \II\!\left[\frac{1}{x},\frac{x-1}{x(1+x)},\frac{1}{x};x\right] \, , \qquad
f^{(E)}_{5}[x] = \II\!\left[\frac{1}{x},\frac{1+x}{x(x-1)},\frac{1}{x};x\right] \, , \\[1em]
f^{(E)}_{6}[x] &= \II\!\left[\frac{x-1}{x(1+x)},\frac{1}{x},\frac{1}{x};x\right] \, , \qquad
f^{(E)}_{7}[x] = \II\!\left[\frac{1+x}{x(x-1)},\frac{1}{x},\frac{1}{x};x\right] \, , \qquad
f^{(E)}_{8}[x] = \II\!\left[\frac{x^{2}-1}{2x(1+x^{2})},\frac{1}{x},\frac{1}{x};x\right] \, , \\[1em]
f^{(E)}_{9}[x] &= -\II\!\left[\frac{x-1}{x(1+x)};x\right]
                 -\II\!\left[\frac{1+x}{x(x-1)};x\right]
                 -2\log 2 \, , \\[1em]
f^{(E)}_{10}[x] &= \sqrt{2}\,\pi^{2}
                  -16\,\II\!\left[\frac{1}{x\,r(x)\,\varpi_{K3}(x)},\frac{(1+x^{2})\,\varpi_{K3}(x)}{x};x\right] \, ,
                  \hspace{6cm} r(x)=\sqrt{-1+34x^{2}-x^{4}}\\[1em]
f^{(E)}_{11}[x] &= \II\!\left[\frac{1}{x},\frac{1}{x},\frac{x-1}{x(1+x)};x\right]
                  +\II\!\left[\frac{1}{x},\frac{1}{x},\frac{1+x}{x(x-1)};x\right]
                  +2\,\II\!\left[\frac{1}{x},\frac{1}{x};x\right]\log 2 \, , \\[1em]
f^{(E)}_{12}[x] &= 32\,\II\!\left[\frac{1}{x},\frac{x^{2}-1}{2x(1+x^{2})},\frac{1}{x};x\right]
                  +9\,\II\!\left[\frac{1}{x},\frac{1-x^{2}}{x(1-x+x^{2})},\frac{1}{x};x\right]
                  -9\,\II\!\left[\frac{1}{x},\frac{1-x^{2}}{x(1+x+x^{2})},\frac{1}{x};x\right] \, , \\[1em]
f^{(E)}_{13}[x] &= \frac{\pi^{2}(1+x^{2})\bigl(2x\,\varpi_{K3}(x) + (x^{2}-1)\,\varpi'_{K3}(x)\bigr)}{2\sqrt{2}\,x\,\varpi_{K3}(x)}
                  -\frac{4(1+x^{2})\,\II\!\left[\frac{1}{x\,r(x)\,\varpi_{K3}(x)},\frac{(1+x^{2})\,\varpi_{K3}(x)}{x};x\right]\bigl(2x\,\varpi_{K3}(x) + (x^{2}-1)\,\varpi'_{K3}(x)\bigr)}{x\,\varpi_{K3}(x)} \\[2em]
\hline\\[1em]
f^{(O)}_{1}[x] &= \II\!\left[\frac{1}{x};x\right] \, , \qquad
f^{(O)}_{2}[x] = \pi^{2}\,\II\!\left[\frac{1}{x};x\right] \, , \qquad
f^{(O)}_{3}[x] = \II\!\left[\frac{1}{x},\frac{1}{x},\frac{1}{x};x\right] \, , \qquad
f^{(O)}_{4}[x] = \II\!\left[\frac{x-1}{x(1+x)},\frac{1}{x};x\right] \, , \qquad
f^{(O)}_{5}[x] = \II\!\left[\frac{1+x}{x(x-1)},\frac{1}{x};x\right] \, , \\[1em]
f^{(O)}_{6}[x] &= \II\!\left[\frac{x^{2}-1}{2x(1+x^{2})},\frac{1}{x};x\right] \, , \qquad
f^{(O)}_{7}[x] = \II\!\left[\frac{1-x^{2}}{x(1+x+x^{2})},\frac{1}{x};x\right] \, , \qquad
f^{(O)}_{8}[x] = \frac{\II\!\left[\frac{(1+x^{2})\,\varpi_{K3}(x)}{x};x\right]}{(1+x^{2})\,\varpi_{K3}(x)} \, , \qquad
f^{(O)}_{9}[x] = \II\!\left[\frac{1-x^{2}}{x(1-x+x^{2})},\frac{1}{x};x\right] \, , \\[1em]
f^{(O)}_{10}[x] &= \II\!\left[\frac{x-1}{x(1+x)},\frac{x-1}{x(1+x)},\frac{1}{x};x\right] \, , \qquad
f^{(O)}_{11}[x] = \II\!\left[\frac{x-1}{x(1+x)},\frac{1+x}{x(x-1)},\frac{1}{x};x\right] \, , \qquad
f^{(O)}_{12}[x] = \pi^{2}\,\II\!\left[\frac{1}{1+x^{2}};x\right]
                  -8\,\II\!\left[\frac{1}{1+x^{2}},\frac{1}{x},\frac{1}{x};x\right] \, , \\[1em]
f^{(O)}_{13}[x] &= \II\!\left[\frac{1}{x},\frac{x-1}{x(1+x)};x\right]
                  +\II\!\left[\frac{1}{x},\frac{1+x}{x(x-1)};x\right]
                  +2\,\II\!\left[\frac{1}{x};x\right]\log 2 \, , \\[1em]
f^{(O)}_{14}[x] &= 8\,\II\!\left[\frac{x-1}{x(1+x)},\frac{x^{2}-1}{2x(1+x^{2})},\frac{1}{x};x\right]
                  -3\,\II\!\left[\frac{x-1}{x(1+x)},\frac{1-x^{2}}{x(1+x+x^{2})},\frac{1}{x};x\right] \, , \\[1em]
f^{(O)}_{15}[x] &= \II\!\left[\frac{x^{2}-1}{2x(1+x^{2})},\frac{1-x^{2}}{x(1-x+x^{2})},\frac{1}{x};x\right]
                  +\II\!\left[\frac{x^{2}-1}{2x(1+x^{2})},\frac{1-x^{2}}{x(1+x+x^{2})},\frac{1}{x};x\right] \, , \\[1em]
f^{(O)}_{16}[x] &= \II\!\left[\frac{x-1}{x(1+x)},\frac{1}{x},\frac{x-1}{x(1+x)};x\right]
                  +\II\!\left[\frac{x-1}{x(1+x)},\frac{1}{x},\frac{1+x}{x(x-1)};x\right]
                  +2\,\II\!\left[\frac{x-1}{x(1+x)},\frac{1}{x};x\right]\log 2 \, , \\[1em]
f^{(O)}_{17}[x] &= -\II\!\left[\frac{1+x}{x(x-1)},\frac{1}{x},\frac{x-1}{x(1+x)};x\right]
                  -\II\!\left[\frac{1+x}{x(x-1)},\frac{1}{x},\frac{1+x}{x(x-1)};x\right]
                  -2\,\II\!\left[\frac{1+x}{x(x-1)},\frac{1}{x};x\right]\log 2 \, , \\[1em]
f^{(O)}_{18}[x] &= \II\!\left[\frac{x^{2}-1}{2x(1+x^{2})},\frac{1}{x},\frac{x-1}{x(1+x)};x\right]
                  +\II\!\left[\frac{x^{2}-1}{2x(1+x^{2})},\frac{1}{x},\frac{1+x}{x(x-1)};x\right]
                  +2\,\II\!\left[\frac{x^{2}-1}{2x(1+x^{2})},\frac{1}{x};x\right]\log 2 \, , \\[1em]
f^{(O)}_{19}[x] &= \II\!\left[\frac{x^{2}-1}{2x(1+x^{2})},\frac{x-1}{x(1+x)},\frac{1}{x};x\right]
                  +\II\!\left[\frac{x^{2}-1}{2x(1+x^{2})},\frac{1+x}{x(x-1)},\frac{1}{x};x\right]
                  -4\,\II\!\left[\frac{x^{2}-1}{2x(1+x^{2})},\frac{x^{2}-1}{2x(1+x^{2})},\frac{1}{x};x\right] \, , \\[1em]
f^{(O)}_{20}[x] &= \sqrt{2}\,\pi^{2}(1+x^{2})\,\II\!\left[\frac{1}{x\,r(x)\,\varpi_{K3}(x)};x\right]\varpi_{K3}(x)
                  -16(1+x^{2})\,\II\!\left[\frac{1}{x\,r(x)\,\varpi_{K3}(x)},\frac{1}{x\,r(x)\,\varpi_{K3}(x)},\frac{(1+x^{2})\,\varpi_{K3}(x)}{x};x\right]\varpi_{K3}(x) \, , \\[1em]
f^{(O)}_{21}[x] &= 2\,\II\!\left[\frac{1+x}{x(x-1)},\frac{x-1}{x(1+x)},\frac{1}{x};x\right]
                  -8\,\II\!\left[\frac{1+x}{x(x-1)},\frac{x^{2}-1}{2x(1+x^{2})},\frac{1}{x};x\right]
                  -3\,\II\!\left[\frac{1+x}{x(x-1)},\frac{1-x^{2}}{x(1-x+x^{2})},\frac{1}{x};x\right] \, , \\[1em]
f^{(O)}_{22}[x] &= \frac{\pi^{2}(1+x^{2})^{2}\,\II\!\left[\frac{1}{x\,r(x)\,\varpi_{K3}(x)};x\right]\bigl(2x\,\varpi_{K3}(x)+(x^{2}-1)\,\varpi'_{K3}(x)\bigr)}{2\sqrt{2}\,x} \\[1em] & \qquad\qquad
                  -\frac{4(1+x^{2})^{2}\,\II\!\left[\frac{1}{x\,r(x)\,\varpi_{K3}(x)},\frac{1}{x\,r(x)\,\varpi_{K3}(x)},\frac{(1+x^{2})\,\varpi_{K3}(x)}{x};x\right]\bigl(2x\,\varpi_{K3}(x)+(x^{2}-1)\,\varpi'_{K3}(x)\bigr)}{x} \, , \\[1em]
f^{(O)}_{23}[x] &= \frac{\pi^{2}(1+x^{2})^{3}\,\II\!\left[\frac{1}{x\,r(x)\,\varpi_{K3}(x)};x\right]\bigl(2x\,\varpi_{K3}(x)+(x^{2}-1)\,\varpi'_{K3}(x)\bigr)^{2}}{8\sqrt{2}\,x^{2}\,\varpi_{K3}(x)}\\[1em] & \qquad\qquad
                  -\frac{(1+x^{2})^{3}\,\II\!\left[\frac{1}{x\,r(x)\,\varpi_{K3}(x)},\frac{1}{x\,r(x)\,\varpi_{K3}(x)},\frac{(1+x^{2})\,\varpi_{K3}(x)}{x};x\right]\bigl(2x\,\varpi_{K3}(x)+(x^{2}-1)\,\varpi'_{K3}(x)\bigr)^{2}}{x^{2}\,\varpi_{K3}(x)}
                  \, .
}}
\\
		\hline
	\end{tabular}
    \caption{The 13+23 uniform transcendentality basis functions  of the 5PM-2SF scattering angle
     $\theta^{(5,2)}
  =
  \sum_{k}
  c^{(E/O)}_k(\gamma) f^{(E/O)}_k(\gamma)$ of \eqn{thetadef}. The functions $f^{(E/O)}_{n}(\gamma)$ are split into even and odd parity under $v\to -v$ and expressed as iterated integrals as in \eqn{IIdef}.
    }
	\label{table:Bfunctions}
\end{table*}

\begin{table*}[h!]
	\setlength{\tabcolsep}{1pt} % Default value: 6pt
	\renewcommand{\arraystretch}{3}	
	\begin{tabular}{|c|}
		\hline
		\scalebox{0.74}{\tabeq{23cm}{\\[-0.3cm]
c^{(E)}_{1}(\gamma) &=
\frac{20\,c_M\,(8-5\gamma^{2})^{2}\,(-1+2\gamma^{2})^{3}}{9\,(-1+\gamma^{2})^{4}}\\ &\hspace{6.2em}
+\frac{1}{2980454400\,(-1+\gamma)^{4}\,\gamma^{7}\,(1+\gamma)^{5}\,(9-37\gamma^{2}+4\gamma^{4})^{6}}\Bigl(
-4355823083673600-1179702085161600\gamma-54109552283740800\gamma^{2} \\
&\hspace{6.2em}
-159983002122451200\gamma^{3}+3069427165327353600\gamma^{4}+4686891227271237600\gamma^{5}-47042025575638816800\gamma^{6} \\
&\hspace{6.2em}
-56734830262509100776\gamma^{7}+412409479034553288984\gamma^{8}+397180727744928503922\gamma^{9}-2026131044951921805198\gamma^{10} \\
&\hspace{6.2em}
-1759665176874660798171\gamma^{11}+1384796864876229493029\gamma^{12}+3217702806553419391752\gamma^{13}+53119928048669241819912\gamma^{14} \\
&\hspace{6.2em}
+23167258285889638974060\gamma^{15}-387462725385751201924500\gamma^{16}-240507171685168749077052\gamma^{17}+1416918939226630113166788\gamma^{18} \\
&\hspace{6.2em}
+1084708124172410230212286\gamma^{19}-3103488656087814090299714\gamma^{20}-2881014492056017195805216\gamma^{21}+4156244755757115442556384\gamma^{22} \\
&\hspace{6.2em}
+4710286363228773565474404\gamma^{23}-3287313630404518266937756\gamma^{24}-4726247055771013355086950\gamma^{25}+1409422282433464742341530\gamma^{26} \\
&\hspace{6.2em}
+2906838846149559548460549\gamma^{27}-239766026295471555846651\gamma^{28}-1123047705264757737820824\gamma^{29}-55188823668667474535064\gamma^{30} \\
&\hspace{6.2em}
+282150534510135862916016\gamma^{31}+44865804833926980432816\gamma^{32}-47237766987118374175488\gamma^{33}-13358330124220216197888\gamma^{34} \\
&\hspace{6.2em}
+5308998157847158352640\gamma^{35}+2358749970257877953280\gamma^{36}-394954827520565401600\gamma^{37}-262814655706225530880\gamma^{38} \\
&\hspace{6.2em}
+18593532328973914112\gamma^{39}+18104686970826084352\gamma^{40}-499018489097355264\gamma^{41}-704323538244337664\gamma^{42}+5768828273295360\gamma^{43} \\
&\hspace{6.2em}
+11840709438996480\gamma^{44}
\Bigr)\, ,\\[0.6em]
c^{(E)}_{2}(\gamma) &=
-\frac{1}{227082240\,\gamma^{8}\,(-1+\gamma^{2})^{4}}\Bigl(
465696000-3667910400\gamma^{2}+16238376000\gamma^{4}-57010504320\gamma^{6}+293689020260\gamma^{8}
+10566697574400\gamma^{9}+774048860236\gamma^{10}\\
&\hspace{6.2em}
-18601728245760\gamma^{11}+475434870121\gamma^{12}
+3388427919360\gamma^{13}-428007546182\gamma^{14}+6883341189120\gamma^{15}-1084682889107\gamma^{16}\\ & \hspace{6.2em} 
-2402832875520\gamma^{17}+1316642357248\gamma^{18}+210386288640\gamma^{19}-714843619328\gamma^{20}
-44291850240\gamma^{21}+146834194432\gamma^{22}
\Bigr)\, ,\\[0.6em]
c^{(E)}_{3}(\gamma) &=
\frac{40\,c_M\,\gamma^{2}\,(3-2\gamma^{2})^{2}\,(-1+2\gamma^{2})^{3}}{(-1+\gamma^{2})^{5}}\\
&\hspace{6.2em}
+\frac{1}{28385280\,\gamma^{8}\,(-1+\gamma^{2})^{5}}\Bigl(
279417600-1490227200\gamma-408038400\gamma^{2}+9792921600\gamma^{3}-1030075200\gamma^{4}\\
&\hspace{6.2em}
-44725739520\gamma^{5}+12475995840\gamma^{6}+182763356160\gamma^{7}-103942382052\gamma^{8}-734460604416\gamma^{9}
+3132824006424\gamma^{10}-7984221020160\gamma^{11}\\
&\hspace{6.2em}
-9891823504669\gamma^{12}
+8273046921216\gamma^{13}+11266287964847\gamma^{14}+2309415026688\gamma^{15}-6002357319091\gamma^{16}-4425682010112\gamma^{17}\\
&\hspace{6.2em}
+3292120340589\gamma^{18}+1440223690752\gamma^{19}-2776010588160\gamma^{20}
-335510765568\gamma^{21}+1292516720640\gamma^{22}+66437775360\gamma^{23}
\\
&\hspace{6.2em}
-220251291648\gamma^{24}
\Bigr)\, ,\\[0.6em]
c^{(E)}_{4}(\gamma) &=
\frac{\gamma\,(12450+58701\gamma+75166\gamma^{2}+54433\gamma^{3}-36150\gamma^{4}-77615\gamma^{5}-35478\gamma^{6}-3705\gamma^{7}+14988\gamma^{8}+9242\gamma^{9})}{2\,(-1+\gamma^{2})^{4}}\, ,\\[0.6em]
c^{(E)}_{5}(\gamma) &=
\frac{\gamma\,(12342+58035\gamma+76426\gamma^{2}+58207\gamma^{3}-40722\gamma^{4}-86225\gamma^{5}-30258\gamma^{6}+3705\gamma^{7}+13188\gamma^{8}+7142\gamma^{9})}{2\,(-1+\gamma^{2})^{4}}\, ,\\[0.6em]
c^{(E)}_{6}(\gamma) &=
\frac{2\gamma\,(27-819\gamma+3141\gamma^{2}-7263\gamma^{3}+1143\gamma^{4}+10705\gamma^{5}-5913\gamma^{6}-3705\gamma^{7}+2498\gamma^{8}+26\gamma^{9})}{(-1+\gamma^{2})^{4}}\, ,\\[0.6em]
c^{(E)}_{7}(\gamma) &=
-\frac{2\gamma\,(-27+1152\gamma-3141\gamma^{2}+5376\gamma^{3}-1143\gamma^{4}-6400\gamma^{5}+5913\gamma^{6}-2498\gamma^{8}+1024\gamma^{9})}{(-1+\gamma^{2})^{4}}\, ,\\[0.6em]
c^{(E)}_{8}(\gamma) &=
-\frac{4\gamma\,(-12234-3123\gamma-70774\gamma^{2}+10401\gamma^{3}+45294\gamma^{4}-2351\gamma^{5}+15822\gamma^{6}-3705\gamma^{7}-7292\gamma^{8}+1050\gamma^{9})}{(-1+\gamma^{2})^{4}}\, ,\\[0.6em]
c^{(E)}_{9}(\gamma) &=
-\frac{1}{88200\,\gamma^{8}\,(-1+\gamma^{2})^{3}}\Bigl(
385875-771750\gamma-1837500\gamma^{2}+4042500\gamma^{3}+7188300\gamma^{4}-17497900\gamma^{5}
-21241500\gamma^{6}+69893600\gamma^{7}+752357266\gamma^{8}\\
&\hspace{6.2em}
+2724435770\gamma^{9}+4066583215\gamma^{10}+204477980\gamma^{11}-3666321886\gamma^{12}-1976705080\gamma^{13}-558224042\gamma^{14}
+287794640\gamma^{15}+318633472\gamma^{16}\\
&\hspace{6.2em}
-3010560\gamma^{17}
\Bigr)\, ,\\[0.6em]
c^{(E)}_{10}(\gamma) &=
-\frac{1}{135\,(9-\gamma^{2})^{15/2}\,(-1+\gamma^{2})^{4}}\Bigl(
8\gamma\,(100010169369-1065466942146\gamma^{2}+289379234907\gamma^{4}+506567453742\gamma^{6}-401937710301\gamma^{8}\\
&\hspace{6.2em}
+137327274588\gamma^{10}-27325236891\gamma^{12}+3441459262\gamma^{14}-279127708\gamma^{16}+14176194\gamma^{18}-411296\gamma^{20}+5240\gamma^{22})
\Bigr)\, ,\\[0.6em]
c^{(E)}_{11}(\gamma) &=
\frac{512\gamma\,(24+60\gamma+137\gamma^{2}+35\gamma^{3}-84\gamma^{4}-59\gamma^{5}-36\gamma^{6}+16\gamma^{8}+4\gamma^{9})}{(-1+\gamma^{2})^{4}}\, ,\\[0.6em]
c^{(E)}_{12}(\gamma) &=
-\frac{9\gamma\,(1-5\gamma^{2})^{2}\,(-3+2\gamma^{2})}{4\,(-1+\gamma^{2})^{3}}\, ,\\[0.6em]
c^{(E)}_{13}(\gamma) &=
-\frac{1}{135\,\gamma\,(9-\gamma^{2})^{13/2}\,(-1+\gamma^{2})^{4}}\,
4\Bigl(
-71441023140-74590066944\gamma^{2}+500006049795\gamma^{4}-430621956207\gamma^{6}
+163543923450\gamma^{8}-32928266424\gamma^{10}\\
&\hspace{6.2em}
+3425665911\gamma^{12}-94322815\gamma^{14}-18483752\gamma^{16}+2293470\gamma^{18}-108424\gamma^{20}+1960\gamma^{22}
\Bigr)\, .
}}
\\
		\hline
	\end{tabular}
    \caption{The 13 even coefficient polynomials of the 5PM-2SF scattering angle $\theta^{(5,2)}$ of \eqn{thetadef}.
   Our $\gamma$-3 prescription yields $c_{M}=1$. }
	\label{table:Cpols}
\end{table*}

\begin{table*}[h!]
	\setlength{\tabcolsep}{1pt} % Default value: 6pt
	\renewcommand{\arraystretch}{3}
	\begin{tabular}{|c|}
		\hline
		\scalebox{0.74}{\tabeq{23cm}{\\[-0.3cm]
c^{(O)}_{1}(\gamma) &=
\frac{40\,c_M\,\gamma\,(-1+2\gamma^{2})^{3}\,(24-31\gamma^{2}+10\gamma^{4})}{3\,(-1+\gamma^{2})^{9/2}}\\
&\hspace{6.2em}
-\frac{1}{2980454400\,\gamma^{8}\,(-1+\gamma^{2})^{9/2}\,(-1+4\gamma^{2})^{7}}\Bigl(
-43961702400-29338848000\gamma+1800773990400\gamma^{2}+1113013440000\gamma^{3}\\
&\hspace{6.2em}
-33063919257600\gamma^{4}-19187055518400\gamma^{5}+370780200806400\gamma^{6}+200398031064000\gamma^{7}-3204383588904960\gamma^{8}\\
&\hspace{6.2em}
-1350746286267980\gamma^{9}+26009084727521280\gamma^{10}+4935234444588840\gamma^{11}-194777322422538240\gamma^{12}+2802591716822751\gamma^{13}\\
&\hspace{6.2em}
+1158445816002478080\gamma^{14}-136163645494652282\gamma^{15}-4984785460423802880\gamma^{16}+732702788132024163\gamma^{17}\\
&\hspace{6.2em}
+14995576561017507840\gamma^{18}-1949086250413888596\gamma^{19}-31068041194548019200\gamma^{20}+2353233689096200176\gamma^{21}\\
&\hspace{6.2em}
+43766208331672780800\gamma^{22}+754509865324200384\gamma^{23}-41205615363729653760\gamma^{24}-6407166496994817792\gamma^{25}\\
&\hspace{6.2em}
+24718104355924869120\gamma^{26}+9102074262240232448\gamma^{27}-7722555048198144000\gamma^{28}-7157003589760167936\gamma^{29}\\
&\hspace{6.2em}
+335811570900664320\gamma^{30}+3534235392136790016\gamma^{31}+346812263002275840\gamma^{32}-928787502393720832\gamma^{33}\\
&\hspace{6.2em}
-25239976610365440\gamma^{34}+94725675511971840\gamma^{35}
\Bigr)\, ,\\[0.6em]
c^{(O)}_{2}(\gamma) &=
\frac{-199207+57344\gamma-1240416\gamma^{2}+43008\gamma^{3}+189180\gamma^{4}+81920\gamma^{6}}{24\,(-1+\gamma^{2})^{5/2}}\, ,\\[0.6em]
c^{(O)}_{3}(\gamma) &=
\frac{256\,(32+149\gamma-72\gamma^{2}-225\gamma^{3}-1665\gamma^{4}-267\gamma^{5}+2058\gamma^{6}+629\gamma^{7}-84\gamma^{8}-270\gamma^{9}-600\gamma^{10}-48\gamma^{11}+160\gamma^{12}+32\gamma^{13})}{(-1+\gamma^{2})^{11/2}}\, ,\\[0.6em]
c^{(O)}_{4}(\gamma) &=
\frac{1}{73920\,\gamma^{8}\,(-1+\gamma^{2})^{7/2}}\Bigl(
646800-1601600\gamma^{2}+12874400\gamma^{4}+19440960\gamma^{6}-3753241191\gamma^{8}+1916708464\gamma^{9}\\
&\hspace{6.2em}
+9884153792\gamma^{10}+5022367328\gamma^{11}+765269538\gamma^{12}-6496448640\gamma^{13}-8107310376\gamma^{14}+1794181664\gamma^{15}\\
&\hspace{6.2em}
+2580243325\gamma^{16}-1690110576\gamma^{17}-337379328\gamma^{18}+1252786176\gamma^{19}+86507520\gamma^{20}-286785536\gamma^{21}
\Bigr)\, ,\\[0.6em]
c^{(O)}_{5}(\gamma) &=
\frac{1}{73920\,\gamma^{8}\,(-1+\gamma^{2})^{7/2}}\Bigl(
646800-1601600\gamma^{2}+12874400\gamma^{4}+19440960\gamma^{6}-4189795449\gamma^{8}+1718270224\gamma^{9}\\
&\hspace{6.2em}
+5987424832\gamma^{10}+4556745248\gamma^{11}+5061590622\gamma^{12}-5842256640\gamma^{13}-7598149592\gamma^{14}+1777549664\gamma^{15}\\
&\hspace{6.2em}
+2108044675\gamma^{16}-1670706576\gamma^{17}-337379328\gamma^{18}+1252786176\gamma^{19}+86507520\gamma^{20}-286785536\gamma^{21}
\Bigr)\, ,\\[0.6em]
c^{(O)}_{6}(\gamma) &=
\frac{105-155\gamma^{2}+1935\gamma^{4}+5091\gamma^{6}-666260\gamma^{8}+154716\gamma^{10}+689664\gamma^{12}-198856\gamma^{14}}{3\,\gamma^{8}\,(-1+\gamma^{2})^{5/2}}\, ,\\[0.6em]
c^{(O)}_{7}(\gamma) &=
\frac{-3150+15120\gamma^{2}-54810\gamma^{4}+204120\gamma^{6}-575161\gamma^{7}-2776550\gamma^{8}-4328730\gamma^{9}-1783320\gamma^{10}+168595\gamma^{11}-98742\gamma^{12}+409600\gamma^{14}}{140\,\gamma^{7}\,(-1+\gamma^{2})^{5/2}}\, ,\\[0.6em]
c^{(O)}_{8}(\gamma) &=
\frac{64\gamma}{135\,(-9+\gamma^{2})^{7}\,(-1+\gamma^{2})^{7/2}}\Bigl(
-71441023140-74590066944\gamma^{2}+500006049795\gamma^{4}-430621956207\gamma^{6}+163543923450\gamma^{8}\\
&\hspace{6.2em}
-32928266424\gamma^{10}+3425665911\gamma^{12}-94322815\gamma^{14}-18483752\gamma^{16}+2293470\gamma^{18}-108424\gamma^{20}+1960\gamma^{22}
\Bigr)\, ,\\[0.6em]
c^{(O)}_{9}(\gamma) &=
\frac{-3150+15120\gamma^{2}-54810\gamma^{4}+204120\gamma^{6}+575161\gamma^{7}-2776550\gamma^{8}+4328730\gamma^{9}-1783320\gamma^{10}-168595\gamma^{11}-98742\gamma^{12}+409600\gamma^{14}}{140\,\gamma^{7}\,(-1+\gamma^{2})^{5/2}}\, ,\\[0.6em]
c^{(O)}_{10}(\gamma) &=
\frac{3\,(13-9\gamma-155\gamma^{2}-415\gamma^{3}-505\gamma^{4}-55\gamma^{5}+375\gamma^{6}+175\gamma^{7})}{(-1+\gamma^{2})^{5/2}}\, ,\qquad
c^{(O)}_{11}(\gamma) =
-\frac{3\,(40-65\gamma-452\gamma^{2}+785\gamma^{3}+440\gamma^{4}-655\gamma^{5}-300\gamma^{6}+175\gamma^{7})}{(-1+\gamma^{2})^{5/2}}\, ,\\[0.6em]
c^{(O)}_{12}(\gamma) &=
\frac{-40612528-254195160\gamma^{2}+6702790\gamma^{4}+24329925\gamma^{6}}{32768\,(-1+\gamma^{2})^{5/2}}\, ,\\[0.6em]
c^{(O)}_{13}(\gamma) &=
\frac{1}{840\,\gamma^{9}\,(-1+\gamma^{2})^{7/2}}\Bigl(
-3675+7350\gamma+19950\gamma^{2}-43400\gamma^{3}-79800\gamma^{4}+191660\gamma^{5}+246540\gamma^{6}\\
&\hspace{6.2em}
-773640\gamma^{7}-222810\gamma^{8}+11432330\gamma^{9}+25264989\gamma^{10}+52528084\gamma^{11}+38260656\gamma^{12}-27696060\gamma^{13}\\
&\hspace{6.2em}
-48986718\gamma^{14}-14410788\gamma^{15}-81724\gamma^{16}+2988720\gamma^{17}+2732032\gamma^{18}+344064\gamma^{19}
\Bigr)\, ,\\[0.6em]
c^{(O)}_{14}(\gamma) &=
-\frac{3\,(-9+56\gamma+99\gamma^{2}-1200\gamma^{3}-315\gamma^{4}+600\gamma^{5}+225\gamma^{6})}{2\,(-1+\gamma^{2})^{5/2}}\, ,\qquad
c^{(O)}_{15}(\gamma) =
-\frac{360\gamma\,(-37-30\gamma^{2}+15\gamma^{4})}{(-1+\gamma^{2})^{5/2}}\, ,\\[0.6em]
c^{(O)}_{16}(\gamma) &=
\frac{-72-1431\gamma-3816\gamma^{2}-5312\gamma^{3}-5592\gamma^{4}-375\gamma^{5}+5896\gamma^{6}+3150\gamma^{7}}{4\,(-1+\gamma^{2})^{5/2}}\, ,\qquad
c^{(O)}_{17}(\gamma) =
\frac{\gamma\sqrt{-1+\gamma^{2}}\,(987-3621\gamma+3911\gamma^{2}+839\gamma^{3}-3046\gamma^{4}+1050\gamma^{5})}{4\,(-1+\gamma)^{2}\,(1+\gamma)^{3}}\, ,\\[0.6em]
c^{(O)}_{18}(\gamma) &=
-\frac{2\,(4078+1041\gamma+26310\gamma^{2}-2773\gamma^{3}+2442\gamma^{4}-1065\gamma^{5}-3646\gamma^{6}+525\gamma^{7})}{(-1+\gamma^{2})^{5/2}}\, ,\qquad\qquad
c^{(O)}_{19}(\gamma) =
-\frac{4\,(-12553-75933\gamma^{2}+11973\gamma^{4}+4321\gamma^{6})}{(-1+\gamma^{2})^{5/2}}\, ,\\[0.6em]
c^{(O)}_{20}(\gamma) &=
-\frac{8}{135\,\gamma\,(-9+\gamma^{2})^{7}\,(-1+\gamma^{2})^{9/2}}\Bigl(
9051089769+290398415544\gamma^{2}+565663163607\gamma^{4}-809584505478\gamma^{6}+350326745469\gamma^{8}\\
&\hspace{6.2em}
-66127781742\gamma^{10}+3216222549\gamma^{12}+942064462\gamma^{14}-199760578\gamma^{16}+17312214\gamma^{18}-736976\gamma^{20}+12440\gamma^{22}
\Bigr)\, ,\\[0.6em]
c^{(O)}_{21}(\gamma) &=
\frac{3\,(-9-56\gamma+99\gamma^{2}+1200\gamma^{3}-315\gamma^{4}-600\gamma^{5}+225\gamma^{6})}{2\,(-1+\gamma^{2})^{5/2}}\, ,\\[0.6em]
c^{(O)}_{22}(\gamma) &=
-\frac{8}{135\,\gamma\,(-9+\gamma^{2})^{7}\,(-1+\gamma^{2})^{9/2}}\Bigl(
100010169369-1065466942146\gamma^{2}+289379234907\gamma^{4}+506567453742\gamma^{6}-401937710301\gamma^{8}\\
&\hspace{6.2em}
+137327274588\gamma^{10}-27325236891\gamma^{12}+3441459262\gamma^{14}-279127708\gamma^{16}+14176194\gamma^{18}-411296\gamma^{20}+5240\gamma^{22}
\Bigr)\, ,\\[0.6em]
c^{(O)}_{23}(\gamma) &=
\frac{2}{135\,\gamma^{3}\,(-9+\gamma^{2})^{6}\,(-1+\gamma^{2})^{9/2}}\Bigl(
-71441023140-74590066944\gamma^{2}+500006049795\gamma^{4}-430621956207\gamma^{6}+163543923450\gamma^{8}\\
&\hspace{6.2em}
-32928266424\gamma^{10}+3425665911\gamma^{12}-94322815\gamma^{14}-18483752\gamma^{16}+2293470\gamma^{18}-108424\gamma^{20}+1960\gamma^{22}
\Bigr)\, .
}}
\\
		\hline
	\end{tabular}
    \caption{The 23 odd coefficient polynomials of the 5PM-2SF scattering angle $\theta^{(5,2)}$ of \eqn{thetadef}.
    Again $c_{M}=1$ in the $\gamma$-3 prescription.
    }
	\label{table:Cpols2}
\end{table*}

\clearpage

\bibliographystyle{JHEP}
\bibliography{2SF-conservative}

\end{document}